\documentclass[letterpaper,12pt]{article}
\usepackage{pdfpages}
\usepackage {amsmath}
\usepackage{graphicx,hyperref}
\graphicspath{ {diagram/} }

\newtheorem{proposition} {Proposition}
\def\noi{\noindent}
\makeatletter
\def\@xfootnote[#1]{%
  \protected@xdef\@thefnmark{#1}%
  \@footnotemark\@footnotetext}
\makeatother

\begin{document}
\baselineskip=19pt  
\raggedbottom
\oddsidemargin
\evensidemargin
\def\n{\noindent}
\def\n{\noindent}
\def\^{\sp}
\def\f{\footnotesize}
\def\t{\tilde}
\def\pd{\partial}
\def\bdt{{\bar \Delta} \theta}
\def\dt{\Delta \theta}
\def\ts{{\tilde \sigma}}
\def\tt{ {\tilde t}}
\def\bea{\begin{equation}}
\def\eea{\end{equation}}
\def\p{\par}
\def\w{\widehat}
\def\what{\w}
\def\H{\cal H}
\def\S{\cal S}
\def\A{\cal A}
\def\I{\cal I}
\def\J{\cal J}
\def\K{\cal K}
\def\ve{\varepsilon}
\def\k{\kappa}
\def\c{\check}
\def\r{\rm}
\def\pa{\parallel}
\def\b{\buildrel \rm def \over =}
\def\u{\underline}
\def\grad{\nabla}
\def\.{\cdot}
\def\lb{\lbrace}
\def\rb{\rbrace}
\def\e{\exists}
\def\s{\setminus}
\def\implies{\Rightarrow}
\def\eqv{\Leftrightarrow}
\def\to{\rightarrow}
\def\notgeq{\geq \hspace*{-10pt} /}
\def\square{\hfill\hbox{\vrule height .9ex width .8ex depth -.1ex}}
\def\sq{\square}

\thispagestyle{empty}

\begin{center}

{{\Large{Asymmetric networks, clientelism and their impacts: households' access to workfare employment in rural India}}}\footnote[$\star$]{This project is funded by the European Union under the 7th Research Framework Programme (Theme SSH.2011.1), Grant acknowledgement number 290752. Bhattacharya also acknowledges an RIS fund as well as institutional support from University of York. Kar acknowledges additional financial and infrastructural support from Centre for Development Economics, Delhi School of Economics. We thank Thomas Cornelissen for extensive comments and suggestions on multiple drafts. We are also indebted to the comments and suggestions from Ashwini Deshpande, Bhaskar Dutta, Deepti Goel, Matthew Jackson, Dilip Mookherjee, Cheti Nicoletti, Kunal Sen and of several seminar and conference participants. Peter Lynn provided very useful help in our sample design. We acknowledge excellent research assistance from Vinayak Iyer, Amit Kumar, Mamta and Nishtha Sharma. Pramod Dubey and Sunil Kumar diligently monitored the household surveys. Of course, the errors are ours.\\Declaration of interest: none.}

\vspace{0.2in}

{\large{Anindya Bhattacharya}}\footnote[$\dagger$] {Department of Economics and Related Studies, University of York, York YO10 5DD, United Kingdom; anindya.bhattacharya@york.ac.uk} ~~{\large{Anirban Kar}}\footnote[$\ddag$]{Delhi School of Economics, University of Delhi, Delhi, India; anirban@econdse.org}~ ~{\large{Alita Nandi}}\footnote[*]{ISER, University of Essex; anandi@essex.ac.uk}

\vspace{0.2in}

[The substantive content of this version is from March, 2021]

\vspace{0.2in}

\begin{abstract}

In this paper we explore two intertwined issues. First, using primary data we examine the impact of asymmetric networks, built on rich relational information on several spheres of living, on access to workfare employment in rural India. We find that unidirectional relations, as opposed to reciprocal relations, and the concentration of such unidirectional relations increase access to workfare jobs. Further in-depth exploration provides evidence that patron-client relations are responsible for this differential access to such employment for rural households. Complementary to our empirical exercises, we construct and analyse a game-theoretical model supporting our findings.

\end{abstract}

\end{center}

\vspace{0.2in}


\vspace{0.2in}


\newpage

\setcounter{page}{1}

\section{Introduction}

\noindent In this paper we explore two intertwined issues: the impact of social networks on some development outcomes and that of clientelistic rural institutions in developing countries. Following Bardhan and Mookherjee (2017, 2018), we understand clientelism as the practice of making private transfers of perishable goods and benefits by a section of the elites (patrons) to a section of the poor and disadvantaged groups (clients). Unsurprisingly, this is not an act of benevolence, but a means of securing socio-political domination by the patron(s). When facing competition from other sections of elites, a patron secures her clients’ political support through such transfers. Studies in clientelism (see Bardhan and Mookherjee, 2017 for a recent survey) indicate that such an institutional structure generates lop-sided political competition, pro-incumbency and favours directed private transfers at the expense of public goods. In these contexts, this paper mainly addresses the following questions.

First, within rural developing economies with unemployment (as in several parts of India), where publicly provided workfare employment could be of considerable value to poor households, what kind of network connections of households, if any, can affect access to workfare jobs. To examine this we use rich data on interpersonal interactions, collected through a primary survey. We find that unidirectional hierarchical connections, as opposed to reciprocal–friendship-like–connections, and the concentration of such links have a significant positive impact on securing jobs. We further argue, both empirically and theoretically, that these network-related features reflect an underlying patron-client power-structure which is associated with differential access to workfare employment for the rural households. Although our analysis does not establish an unambiguously causal relationship, we demonstrate that the results remain robust to a variety of sensitivity checks.

The main building block of our investigation is data on day-to-day interactions of rural households in the spheres of economy, societal living and political activities. We conducted a household survey in 36 sampled villages in three states of India (resulting in a total of 3454 households) using personal interviews at the household level\footnote{The first phase of the household survey, covering mainly Odisha and Maharashtra, took place during March-April, 2013 and the second phase, covering mainly UP, took place in November-December, 2013. Details of the sample design can be found in Appendix B.}. We gathered information on the items of services each household receives in spheres of: (a) economic interactions – such as whom the household primarily depends on for getting productive inputs, for selling outputs, for working as informal farm-labourers, for informal loans, etc.; (b) in social interactions – such as whom the household primarily approaches for advices on family matters and disputes, religious matters, etc. and (c) political interactions – such as whom, if any, the household accompanies to political events, depends on for accessing governmental bureaucrats, etc.\footnote{Find the comprehensive list in Section 3.} We divide these interactions into two categories. Some households do not depend on any particular person for receiving a service, while others have fixed providers. For instance, within the same village, we found agricultural labourers who take up jobs as these come by, while others are primarily dependent on one or a couple of employers over the years. We conceive the former as purely anonymous market-like transactions. The latter type: i.e., personalized interactions, provide the network data for our analysis. Thus, a network-link in this study does not represent merely whether or not two households interact with each other, but whether such an interaction has a special importance as perceived by the household receiving a service.

We observed huge variations in this network data. There are two kinds of variations which are particularly important for us in this paper. The first link characteristic we consider is the direction of flow of services. Within a village, we found some pairs of households providing service(s) to each other, whereas among some other pairs the flow is unidirectional. Reciprocal and unidirectional links indicate qualitatively different types of interactions. The former hints at equal status like friendship, caste or family ties: households offering insurance and information to each other. In contrast, the latter is indicative of dissimilar/hierarchical positions in a community, like buyer-seller, celebrity-follower, patron-client, etc. This has important implications for economic outcomes. For example, while mutual insurance traded through reciprocal links among small farmers can be effective against idiosyncratic shocks, a unidirectional link between a farmer and a trader is likely to be more effective against correlated shocks\footnote{Note that our concept of reciprocal and unidirectional link is different from that of a directed link. A directed link is commonly understood as a link that can be established unilaterally (does not require consent of the receiver) like clicking on a hyperlink but in our context all interactions naturally require consent from both households.}. Therefore we classify our network linkages into two groups - reciprocal and unidirectional. 

The second feature we consider is the concentration of unidirectional links. Some households were observed to have received multiple services from others. For example, a household $X_1$ may mention household $Y_1$ as a regular employer and also as a person who usually takes a member of household $X_1$ to political events. We consider the interaction between $X_1$ and $Y_1$ as more intense than another one, where a household $X_2$ receives only one kind of service from some $Y_2.$ Concentration is important, because providers of multiple services have stronger bargaining power over the receivers. A similar strand of understanding can be found in the literature on interlinked contracts/markets (see, e.g., Bardhan and Udry, 1999, Chapter 9).

We explore the importance of concentration and direction (reciprocal or unidirectional) of links on access to jobs under the Mahatma Gandhi National Rural Employment Guarantee Scheme (MGNREGS hereafter) in India, the largest workfare scheme in the world which was initiated in 2006. The MGNREGS was supposed to provide a maximum of 100 days of unskilled manual work at a government-stipulated minimum wage to each rural household during our study period (prior to 2013)\footnote{We refer to the official website of this scheme- http://www.nrega.nic.in/netnrega/home.aspx  for details.}. The scheme was planned to be demand-driven. However, it has been observed that in reality there is substantial evidence of job-rationing and targeting under this scheme. Sukhtankar (2017) has summarized some of the research in this regard. Since the village {\em panchayat}\footnote{{\em Panchayat} is the rural local government, the lowest level elected body in India.} is primarily in charge of running the scheme, it can be expected that social network may play an important role in determining the allocation of jobs. 

We take two measures of access to MGNREGS: - (i) whether a household has ever worked under the MGNREGS since its introduction, and (ii) the number of days a household has worked under this scheme in the last 12 months preceding our survey. In our sample, 46\% of households have ever worked under the MGNREGS and the average number of work days in the last 12 months is 14.1. We find that an extra unidirectional link increases a household’s probability of ever working under the MGNREGS by about 28\% of the average. Similarly, it increases the number of workdays by 24\% of the average. By contrast, an extra reciprocal link has no significant impact on either of the indices of access to MGNREGS.
 
As far as the concentration of unidirectional links is concerned, one standard deviation increase in our concentration measure is associated with an increase of 0.06 probability of ever working under the MGNREGS and 1.6 extra work days. We use village fixed effect in all regressions. 

As the second--follow-up--research question, we delve deeper to identify institutional factors that may give rise to the results above. Our theoretical model suggests that clientelism leads to concentration of unidirectional links and differential access to workfare jobs between clients and others. We also observe from our data that households that have reciprocal links are relatively better off than receivers of unidirectional links, while providers of unidirectional links are the richest and are predominantly from socially dominant caste groups. Together these suggest that a patron-client relation may be at work. To examine this, we define a ‘patron’ as a household which is a provider of unidirectional links to at least 5\% of sampled households in a village. A ‘client’ is a household which receives at least one unidirectional link from a patron. Once again, using village fixed-effect regressions, we find that a client household has a 15\% higher probability (than average) of working under the MGNREGS and it gets 31\% more workdays (than average). We acknowledge that, due to sampling, we may have failed to identify some patrons and hence our estimate of the client effect may be biased. However, we demonstrate that our estimates are robust to alternative specifications and we rule out alternative explanations.

For some readers our findings may seem unsurprising: it might look quite natural that there is a link-related relationship driving access to NREGS, particularly given the fact that certain agents, active in the {\em panchayat}, are supposed to control this access (by law). However, this observation cannot explain why reciprocal connections fail to affect access. Further, recall that people controlling the local rural government are political functionaries. But, as we verify below in Section 6, unidirectional connection with patrons having occupations like private business also results in better access to such jobs.

While in the next section we provide details about the several strands of research to which this work contributes, here we contextualize this paper against the background of existing literature in the two main areas mentioned above. The literature on the impacts of socio-economic networks/connections is too huge to permit a full review within the confines of this paper. There have been numerous studies on understanding the impacts of social networks on economic outcomes, particularly in the context of developing countries (e.g., Banerjee et al., 2013, 2019; Topa and Zenou, 2015; Bramouille et al., 2016; Cruz et al., 2017, Bandeira et al., 2020). But the usual ways in which network-related variables are constructed fall into the following main categories. First and foremost, the bulk of the literature looked at reciprocal networks, in other words, the relation captured by the network structure is like friendship or collaboration or exchange of services. Furthermore, studies in networks, while defining links, have not taken into account the possibly rich structure of relationships between people or entities. Even when multifarious aspects of relationships are present in the underlying data, often these get summarized into a single dimension: whether two entities are linked or not (e.g., as in Banerjee et al., 2013). By contrast, our focus is on unidirectional networks and we use the multiple kinds of relations between households present in our data to measure the concentration of links. The second area, identifying and quantifying clientelistic institutions and measuring their impacts, also has thorny issues. For example, household surveys meant to study {\em political} clientelism, directly asking questions on vote buying and political support, are likely to suffer severely from underreporting and misreporting. Thus, works on clientelism to date have primarily relied on indirect evidence.  For instance, Anderson et al. (2015) have used variation in landholding and the population of the dominant caste across villages in Maharashtra (a state in India) to predict when clientelism is more likely to arise. Bardhan and Mookherjee (2012) have relied on a dummy election conducted by the authors (as part of a household survey) to measure political support for the ruling party and relate it to clientelism. Wantchekon (2003) used a field experiment in collaboration with political parties, and so on. Though these papers are extremely valuable to understand the nature and impacts of clientelism, all of these have primarily relied on indirect evidence and proxies of clientelistic practices. We offer a new direct way of identifying the variation of rural clientelistic institutions through the structure of socio-economic networks. In our context two recent papers seem especially worth-discussing. Bandeira et al. (2020) recently analyzed how social ties affect a particular kind of programme delivery in rural Uganda and found that ``delivery agents favor their own social ties over ex-ante identical farmers". But the components of the programme concerned--``training in modern agricultural techniques as well as improved seeds"--are not quite amenable for use as clientelistic instruments and they have not tried to compare, in contrast to us, the impacts of symmetric vis-a-vis unidirectional connections. Maitra et al. (2020) have studied, in context of rural West Bengal (in India), the relative efficacy of providing micro-credit (a suitable instrument for clientelistic influence) via local traders vis-a-vis politically influential delivery agents. Their findings are in line with ours.  

The paper is organized as follows. In the next section we provide some further discussion on our contribution to several sub-areas of the existing literature. Section 3 provides the details about the construction of our networks. We provide a supporting game-theoretical model in Section 4. The section following contains our empirical strategy as well as details of the variables we use. The empirical results are discussed in Section 6. Some concluding remarks are presented in Section 7. The tables have been collected at the end of the main body of the paper: following the list of references cited. At the very end are three appendices. The first one gives the proof of our main theoretical proposition, the second one provides the details of our sample design and the third one collects the Tables.

\section{The existing literature and our contribution}\label{literature}

As we mentioned in the Introduction, this work makes secondary contribution to several streams of research. In these contexts, below we mention the significant related works and remark on what new we introduce to the existing literature.

\medskip \noi {\em{Impact of position/centrality of agents in a network structure and value of connections or important nodes}}

\noi We have already contextualized our work within this stream of literature. In addition, a body of literature exists on the value or impact of connections: very notable among them being Bandeira et al. (2009), Munshi and Rosenzweig (2016) and Markussen and Tarp (2014). Our contribution, naturally, falls also into this terrain of research. We repeat that within this sub-area the novelty of our finding is that it is not that merely that connections matter, but that unidirectional connections, especially with powerful entities, matter. Banerjee et al. (2013) is  especially notable in our context, as it also explores the role of ‘powerful’ nodes. However, the kinds of day-to-day socio-economic relations they took as primitive are more or less ‘reciprocal’ in nature. Thus, links in their study represent ‘friendship’. We repeat that, by contrast, our focus is on ‘hierarchical’ relations. This leads to a further important difference between our work and earlier studies. Measures of positional power for symmetric networks, such as eigenvector centrality, are based on the idea that the centrality/power of a node depends on how central its neighbours are. Once again, the underlying assumption here is that links are reciprocal in nature. Once this assumption is set aside and reciprocal and unidirectional links are considered separately, we find that the unidirectional neighbour of a powerful agent may not be powerful at all (a point noted by Herings et al., 2005 in their theoretical work). In fact, we infer that she is a client of the powerful patron. The consequence of such a relation is ambiguous.

\medskip
\noi {\em{Impact of institutions on development-related outcomes in general}}

\noi While one of our foci is on rural clientelistic institutions, our approach ‘measures’ institutions {\em not in terms of some exogenously given characteristics} but {\em endogenously}--by using data on day-to-day interactions as the primitive. In this respect our work is different from apparently similar works such as that of Acemoglu et al. (2014), which looks into the impact of connection with ‘elites’. ‘Elites’ in their case are historically given. Moreover, unlike, for example, as in Goldstein and Udry (2008), we do not measure the impact of having power only in the sphere of formal politics (more on this, especially in the context of the allocation of MGNREGS jobs, below). We conceptualize the exercise of power (and the reciprocal idea of dependence) as dominance in several aspects of living.

\medskip

\noi {\em{(Mis)Targeting of welfare schemes}}

\noi Our paper also adds to the large body of literature on the (mis)targeting of welfare schemes. Besley et al. (2012) and Markussen (2011) have provided evidence of political distortion in the allocation of below-poverty-line (BPL) cards in India. Platteau (2004) and Pan and Christiaensen (2012) found  similar instances of mistargeting in Africa. In contrast, Alatas et al. (2012) did not find evidence of political capture in the identification of ‘poor families’ in Indonesia. Our paper shows that, apart from the problem of clientelistic distortions, the allocation of MGNREGS work is reasonably well-targetted  towards the poor. For instance, households whose maximum education among all the members is up to the pre-university level are significantly more likely to work under the MGNREGS than those households where at least one member has university-level education. Similarly, households with either salaried members or ownership of business are less likely to work under the MGNREGS compared to agricultural labourers and farmers. A similar pattern is also manifested in terms of land-ownership. Another consistent pattern that emerges from our analysis and suggests possible mistargeting is the following. Muslim households, even after controlling for other possible determinants, are significantly less likely to work under the MGNREGS (and received fewer days of work in the 12 months prior to our survey) compared to `lower caste’ households. See, for example, Tables A4 and A5 in Appendix C for details.

\medskip
\noi {\em{Allocation of MGNREGS jobs}}

\noi Studying a few villages in a district in West Bengal (another state in India), Das (2015) found evidence of positive impact of political clientelism in securing such jobs: households which are politically active and supporters of the local ruling political party are more likely to receive the benefits in terms of participation, number of days of work and earnings from this programme. Dey and Bedi (2015) have reinforced such a finding. By contrast, Chau et al. (2017) have found that all activists, irrespective of their political affiliation, receive higher benefits compared to politically inactive households. A related but conceptually distinct body of literature on a `pork-barrel’ type allocation of the MGNREGS funds is also available. This literature focuses on how political affiliation distorts fund-targeting across administrative units. For instance, Gupta and Mukhopadhyay (2016) have found that more funds are allocated to blocks where the Indian National Congress (a centrist political party) has a lower initial vote share. Dey and Sen (2016) have found a similar feature in MGNREGS fund-targeting from {\em panchayat}-levels to the (lower) village-levels in West Bengal. Our study generalizes such findings: we find such evidence for a larger sample, spreading over three states of India, with quite diverse economic and political histories.

\section{Networks, measures on these and their construction from survey data}\label{network}

\subsection{Survey}\label{survey}

Our analyses here are based on primary data that we collected through household surveys in 2013  from 36 sampled villages across three states of India: Maharashtra, Odisha and eastern part of Uttar Pradesh (UP). One major reason for our choice of these three states is that all three are major states with each having a distinct major language and cultural heritage (details of our sample design is given in Appendix B). Additionally, these states comprised of regions with diverse historical patterns of administrative and land-revenue systems during the British-colonial period (permanent settlement, princely states, taluqdari system, ryotwari system) which have been shown to affect the development of post-colonial public institutions and in turn, economic outcomes. Maharashtra and Odisha also have coastal regions which are expected to have occupational diversity as compared to the more agricultural non-coastal regions. We thus selected the sample using a stratified design to ensure sufficient variation across these features.

As one of our primary objectives was to collect information about interpersonal provision of services (i.e., the basis of network data), we aimed to collect information about the entire village. However, cost considerations necessitated putting a cap on the number of households we could interview (our target sample size was 3600). So, at each village we conducted personal face-to-face interviews, mostly with the head of household (HoH), in all households in a village if the village contained less than 100 households. Otherwise we interviewed upto 110 randomly sampled households in each village which contained more than 100 households. This resulted finally in a sample of 3454 households (1185 in Maharashtra, 1120 in Odisha, 1149 in Uttar Pradesh).

At these interviews we collected basic socio-demographic information (education, marital status, age, gender, relationship to the HoH) about all household members as well as about other individuals considered to belong to the same household but who did not live at these addresses, and female children of these household members who were now married and lived elsewhere. For every household interviewed, we collected information about the main occupation of the household, its wealth, some items of expenditure as well as their experiences with the MGNREGS programme, other welfare programmes, medical services and some details of credit history. Finally, we collected detailed information about economic, social and political interpersonal interactions of each household, the details of which we take up in the next sub-section.

\subsection{Network}\label{network index}

For each household the information about interpersonal interactions are from economic, social and political aspects of lived experience. Our data include five economic services: namely, purchasing production input, taking land on sharecropping or rental contract, selling agricultural output, working as a labourer and getting credit. We also had three political services; accessing welfare schemes ({\em except MGNREGS}), receiving guidance on matters related to political participation/activities and getting help for employment-related dispute resolution. Items of services in the social sphere included receiving guidance on religious matters and family dispute resolution. The household head was asked to name the individual or organization from whom the household {\em primarily} received a service over the years. Households were free to name more than one providers or none as the case may be. For instance, regarding credit, we see three kinds of responses: $(i)$ several lenders, but none of them is primary, $(ii)$ the primary lender is a financial institution like a bank and $(iii)$ primary lender is an individual. In case a primary lender exists and is an individual, we collected detailed information about her location, occupation, caste and social and political position. We also asked whether the interviewee household receives any of the remaining services from the said individual. Therefore for each sampled household, we have data on primary providers of the services. {\em We emphasize that, naturally, in context of this exercise, we excluded ``providing access to MGNREGS jobs" as an item of service.}

Given these items of information, we classify interpersonal interactions into two categories. If an interaction is of a personalized nature where the provider of services is a specific individual, i.e., like type $(iii)$ in the paragraph above, then we define it to be a link {\em from} the receiver {em to} the provider. Instead, if an interaction belongs to one of the former two types - either with an organization or with no specific provider - then we consider it as an impersonal ``market-like" transaction. Interpersonal links form the building blocks of our analysis. Each link represents a transaction of service between one of our sampled household and another individual. Note that there can be multiple links between a pair of receiver and provider. For instance, a receiver $X$ can obtain political advice/directive as well as credit from a provider $Y$. Here $X$ has two links as receiver from $Y$. Since we also ask each receiver whether she provides some services (from our list) to $Y,$ we know about the reciprocal relation, if any, as well.  Note that all providers are not necessarily part of our sample (we did not impose such a restriction). When the provider is part of our sample, we identify her by the household she belongs to. Details about providers also allow us to match them across all sampled households in a village. When combined across households, for each sample village, we obtain a network of interpersonal relationship.

Next we create, for each sampled household, a number of network-related measures/indices.

The {\em first} index classifies the sampled households into three categories based on their status as receivers. These are:\\
$(i)$ Households which do not receive any link: these households obtain any necessary services through market-like transactions;\\
$(ii)$ Receivers with \lq reciprocal links': A household $X$ belongs to this group if it receives at least one service through a personal link. Moreover $X$ reciprocates and provide some service or other to all its providers.\\
$(iii)$ Receivers with \lq unidirectional links': These households must have at least one service received from a personal provider without any reciprocation.\\
This classification scheme, first, distinguishes between those who use personal links and the rest and then separates reciprocal links from unidirectional links. If two households provide service to each other they are likely to be in a horizontal relationship like kinship or friendship. In contrast, unidirectional links are likely to represent vertical relationship between households of unequal status\footnote{Note that group $(i)$ can be further divided into two categories - only service providers and those who are neither receiver nor provider. This distinction is not central to our objective of identifying clients, hence we prefer to club them together. Moreover this is likely to have measurement error because we asked households about services they receive and whether they reciprocate or not. Thus a household who is solely a provider can not be identified unless those who receive services from her belong to our sample. This can lead to mis-categorization of providers as one with no-link.}. Out of 3454 sampled households, 1728 households have not received any links. There are 258 households with only reciprocal links and the rest, 1464, have at least one unidirectional link. Since reciprocal links represent horizontal relationship, we expect households with only reciprocal links are likely to be more well-off compared to households with at least one unidirectional link. This is reflected in our sample (see Table 3.1). Among category $(ii)$ households, for 23.3\% the primary occupation is that of manual labour and 11.6\% are salaried, while in category $(iii)$ 47.8\% have manual labour as the primary occupation and only 2.7\% are salaried. The former also have on average more land (2.01 acres compared to 1.3 acres for the latter) and non-land asset\footnote{Based on our non-land asset index, discussed in subsection 5.2. It is a score out of a maximum of 6.} (2.3 compared to 1.7) than the latter. 

The next network-related index we use is the {\em degree} of unidirectional and reciprocal links for each household. Degree of unidirectional link of household $X$ is the number of households with whom $X$ is unidirectionally linked as a receiver. Similarly degree of reciprocal link of $X$ is the number of households with whom $X$ has reciprocal links. Note that if $X$ belongs to category $(ii)$ above then its degree of unidirectional links is 0.

We construct yet another measure for capturing how intensely a household is dependent on another for services. Consider the following example. Suppose that $X$ receives one unidirectional link each from three distinct providers, while $Y$ receives three from a single provider. It is intuitive that dependence of $Y$ over her single provider is more intense than that of $X$. We capture this by a {\em concentration index} defined as follows. Suppose $d_{XK}$ is the number of unidirectional links that $X$ receives from provider $K$. Then the 
\lq Concentration index' of $X= \sum_{\{K\mid X \text{ receives unidirectional link from } K\}} d_{XK}^2$. We normalize concentration index by taking the \lq z-score'. We also consider a variation {\em weighted concentration index} as follows. \lq Weighted concentration index' of $X= \sum_{\{K\mid X \text{ receives unidirectional link from } K\}} w_{XK} d_{XK}^2$. Here $w_{XK}$ is the number of spheres (out of 3 - economic, political and social) in which $X$ receives unidirectional link(s) from $K$. This index puts higher weightage on interlinkage of spheres. Once again we normalize weighted concentration index by taking \lq z-score'.

One possibility is that owing to sampling in several villages, what we identify as a unidirectional connection might be, in reality, part of a triadic connection--involving three entities--which we are failing to observe. This consideration is important as triads have been found to be effective in initiating and sustaining socio-economic transactions in rural economies (e.g., Chandrashekhar and Jackson, 2018). We did explore into this possibility. However, with the types of links we have worked with we did not get any triad even for the villages where we interviewed {\em all} the households.

Our final measure uses the idea of star-like networks to identify \lq important providers' in a village. A person who provides unidirectional services to sufficiently many households in our sample is defined as a \lq patron'. We take 5\% of sampled household in a village as the cutoff. A household that receives unidirectional link from a patron is called a \lq client'. All households in our sample (including those without links) which are neither a client nor a patron, are called \lq non-clients'. We also compute a village level \lq clientelism score'. For each village, it is the weighted sum of unidirectional links provided by patrons to all their clients weighted by the number of their clients{\footnote{Here we express this score precisely using symbols. Let $E$ be the set of patrons in a village (for some villages this set may turn out to be empty). For a patron $j \in E,$ let the number of her
clients be $c_j$ and the total number of unidirectional links provided by this patron $j$ be $n_j.$ Then the clientelism score of this village $C$ is $\sum_{j \in E} c_j n_j.$}}.

It is possible that due to sampling, some patrons remain unidentified. Recall that for each village having more than 100 households, we surveyed upto 110 sampled households only. Further recall that our respondents were free to mention, for each kind of service, primary provider(s), if any, and {\em not necessarily residing in the same village.} Thus, for quite a few households residing in a sample village A may have a service-provider located in a neighbouring village B or even in a nearby town C. These factors introduce error in our measurement of patrons.{\footnote{In 2014 we undertook a household survey of several such {\em patrons} (see Bhattacharya et al., 2018) and in course of that survey we indeed found evidence that some providers for a large village, who would have been coded as patrons had we had a census of the village, have not been coded as such as per our criterion in this paper.}} As a result some clients could also be misidentified as non-clients. Despite this measurement error, we show in Section \ref{other} that our results are quite robust. Summary statistics (see Table 3.1 and 3.2) also show that, as expected, clients are relatively poorer than non-clients. Out of 2850 non-clients, 31\% are manual labourer and 8.6\% are salaried. Corresponding figures for 591 clients are, 41.8\% and 1.5\% respectively. Similarly, households with above median concentration score is relatively poorer than those below it.

\section{A theoretical model}\label{model}
This section provides a theoretical framework supporting our empirical analysis. We construct a parsimonious model to illustrate that clientelism gives rise to concentration of links and star network structure in social interactions. Our model also predicts that clients get more workfare jobs (MGNREGS in our empirical section) than non-clients.

There are two resources and two time period in our model economy. As discussed in the previous sections, these resources can be economic such as access to cultivable land or they can be socio-political such as access to dispute resolution mechanisms. Here we abstract away from specificities of different resources and call them $r_1$ and $r_2$. Agents are unequally endowed in terms of access to resources. There is only one agent, named agent $0$, who is endowed with access to both the resources. There are two more agents, who have endowment of one resource each. Agent $1$ endows access to only $r_1$ and $2$ endows access to only $r_2$.  The remaining $n$ agents, $P=\{3,\ldots, n+2\}$, do not have any of the resources. Since $0$, $1$ and $2$ control access they will be called `elites' and $P$ is the set of the rest \lq poor'. 

Poor agents must consume $r_1$ and $r_2$ in both periods. They can access the resources through following two channels. First, through market transactions like formal land market or legal recourse for dispute resolution. Alternatively one can get access through personal link with an elite. For example, a link with either $0$ or $1$ allows a poor to obtain $r_1$. A link with $0$ gives access to both $r_1$ and $r_2$. Resources and associated benefits are identical across all channels. For instance, quality of dispute resolution remains the same irrespective of whether it was obtained through a personal link or at formal institutions. Therefore, for accessing resources, formal transactions and access links are perfect substitutes of each other. However, these have different costs. Formal market transactions entail a search cost (or transaction cost). An agent $k$ pays search cost $s_k$ for each resource in each period. For each $k$, $s_k$ is exogenous and common knowledge. We assume that search costs are evenly spaced between $0$ and $1:$ that is, $s_3=\frac{1}{n},s_4=\frac{2}{n},\ldots,s_{n+2}=1$. On the other hand, a personal link with elite $0$ costs $c$, while link with $1$ and $2$ are costless. We assume that $c$ is strictly higher than search costs, that is $c>1$. Link formation takes place at the beginning of first period and once formed a link lasts for two periods, subject to consent of the elite at the beginning of the second period. Poor pay the link formation cost, if any, just once, at the time of link formation. Further we assume that access through links are non-rival: that is, several agents can access a resource through a single elite without any loss in benefit. Elites do not directly benefit from a link.

Let us first look at the benchmark case. Here, an elite does not have any strategic choice. She is indifferent between forming and not forming a link. Since links to $1$ and $2$ are costless, all poor establish these two links for accessing resources. In our game protocol, to follow, a poor can have at most one personal link. We make this assumption to capture clientelistic loyalties. Just for the sake of comparison, even if we impose the same on the benchmark, the outcome does not change substantially. A poor now establishes one link with either $1$ or $2$ and obtain the other resource from market. Thus in the benchmark case, resources are accessed through separate channels. Hence concentration index of the equilibrium network is 0. If we bring in replicas of $1$ and $2$ - then the equilibrium network ceases to be star-shaped as well. Therefore link concentration and star shape do not arise naturally in this economy.

Let us now introduce the source of possible patron-client relation emerging in our model. We assume that elites aspire political office. To this end, an elite may use promise of workfare jobs along with access to $r_1$ or $r_2$ for inducing voters to vote for her. Such jobs are state funded but the elected official has discretion over their allocation. An election takes place at the end of the first period. We assume that public work is available only in the final period, after the election. Maximum amount of per capita job (as in the case of MGNREGS) is exogenous and is denoted by $b$. All elites, $0,1$ and $2$ contest in the election. There is also a fourth candidate, a non-native, denoted by $N$, whose role will be explained below. Only poor vote in the election. To keep our analysis simple, we model the election as a random draw among candidates, where each candidate's probability of selection is equal to proportion of vote obtained. Thus, if a candidate receives $k$'s vote then her winning probability increases by $\frac{1}{n}$ irrespective of how others vote.

Timeline of our model is as follows. In period 1, first, each elite chooses a subset of poor, who get the consent to form link with her. Let us denote these sets by $P_0$, $P_1$ and $P_2$. All agents in $P_0$ have consent to form link with $0$ and so on. Next, a poor chooses exactly one elite from offers he has received and form a link with her. Otherwise, he can also choose to remain unlinked. Thus a poor can have at most one link. All agents in $P$ make this decision simultaneously. If $k\in P$ establishes a link with an elite $l$, we call $l$ the patron of client $k$\footnote{This definition is restricted to theory section only and should not be confused with the index of patron and client given in Section $\ref{network index}$.}. Once links are formed, poor consume first period resources and depending on link status, pay appropriate costs. Before the end of first period, clients are offered a contract (to be discussed next) for period 2 by their respective patrons. At the end of period 1, the election is held. Each poor agent decides whom to support during election. Since secret ballot is used, a poor person may choose not to support his patron without possibility of detection. In period 2, all contracts are executed\footnote {We assume that contracts are not reneged at this stage.}. Finally, period 2 resources are consumed and depending on the current link status appropriate costs are paid. No new link can be formed in period 2.

Due to secret ballot, patron cannot directly control her client's vote. Instead she uses a contract to induce a client to vote in her favour. A typical contract uses two instruments:  $(i)$ threat that consent for link will be withdrawn in period 2 in case the patron is not elected; and $(ii)$ reward of public work if the patron is elected\footnote{Note that contracts can only be contingent on verifiable events, which is winning or losing in this case.}. This contract represents patron-client relation, offer of personalized resources in exchange of political support, in our model. We further assume that all elites are inefficient and can only deliver $\theta b$ (where $\theta<1$) amount of public work. The non-native candidate, $N$, embodies a non-clientelistic option in the election. She delivers the full quota of public work $b$ and gives it to all poor irrespective of their link status (hence it is non-clientelistic).  The non-native candidate, however, does not have any strategic choice in our model.

Payoffs are as follows. An elite, if elected, incurs an effort cost for providing public work. Expected payoff of an elite $l$ is: $\frac{s_l}{n} \left [R- em_l(\theta b) \right ]$. Number of agents who votes for $l$ and number of $l$'s clients are denoted by $s_l$ and $m_l$ respectively. As voting is a strategic choice, $m_l$ is not necessarily the same as $s_l$. The total contracted amount of public work, in case $l$ wins, is $m_l(\theta b) $. Ego rent from office is denoted by $R$ and per unit effort cost is denoted by $e$. We assume that $R$ is sufficiently large\footnote{$R\geq 2ne(\theta b)$ ensures $s_l \left [R- em_l(\theta b) \right ]\geq (s_l-1) \left [R- e(m_l-1)(\theta b) \right ]$} so that $l$ always wants an extra client as long as his support is guaranteed. On the other hand $l$ does not want a client who is sure to vote against her because it increases cost without a compensating increase in win probability. A poor maximizes his expected payoff from workfare job and access costs across two periods. Since resources are identical across access channels, benefit from them are omitted. Agents do not discount future benefits.

Denote the following subsets of $P$: $\Pi_0=\left \{k\in P\mid s_k\geq \frac{b(1-\theta)}{2}\right \}$; $\Pi_1=\left \{k\in P\mid s_k\geq b(1-\theta)\right \}$ and $\Pi_U=P\setminus \Pi_0$

We impose a couple of parameter restrictions: $(a1)$ $b\geq 3$; $(a2)$ $1<c<\min\{2b(1-\theta),2\}$ and $(a3)$ $\frac{1}{n}<\left [min\left \{1,b(1-\theta)\right \}-\max\left \{\frac{c}{2},\frac{b(1-\theta)}{2}\right \}\right ]$. The assumption $(a1)$ puts a lower bound on per capita public work, if $b$ is too low then it can not be used to induce voters. The assumption $(a2)$ imposes an upper bound on link cost. If $c$ is too high then link with $0$ will not be used at all. Restriction $(a3)$ ensures that population is not distributed too sparsely. We state the main result here and provide some intuition. A proof can be found in Appendix A.
\begin{proposition}\label{theory}:
$(a)$ The following is a subgame perfect Nash equilibrium of this game:\\
$(i)$ Each $i\in \Pi_0$ receives an offer from elite $0$. Each $i\in \Pi_1$ receive offers from elite $1$ and $2$ as well. $(ii)$ Each $i\in \Pi_0$, if offered link with $0$, will form the link. Each $i\in \Pi_U$ remains unlinked. $(iii)$ Each $i\in \Pi_0$, who has formed link with $0$, votes for $0$ and each $i\in \Pi_U$ votes for $N$.\\
$(b)$ Clients get more public work than non-clients. \\
$(c)$ If $b(1-\theta)\geq \frac{6}{7}$ then the above is the only equilibrium of this game.
\end{proposition}
Recall that link concentration and star-shaped network do not arise in our benchmark model. However, prediction changes dramatically when the elites contest for the political office. Despite high link cost with $0$, all those whose search costs are above $\frac{b(1-\theta)}{2}$, choose to access both the resources through $0$. This generate link concentration as well as star-shaped network. Creating replica of $1$ and $2$ do not change the equilibrium network structure. The underlying forces are as follows. We know that elites use conditional contracts to gain client's loyalty. Here, multiple resources owned by $0$ give her an advantage over $1$ and $2$. $1$ and $2$ own access to one resource each and hence have weaker threat compared to $0$. This means $1$ and $2$ can only rely on those with search cost higher than $b(1-\theta)$. Whereas $0$ can rely on a superset of poor, those with search cost higher than $\frac{b(1-\theta)}{2}$. Next, note that strategic interactions at the election stage constitute a coordination game. It is profitable for a poor to support a candidate with large support base because it increases his chance of obtaining public work. If potential client strength of $1$ and $2$, $[1-b(1-\theta)]$, is sufficiently small then all of them prefer to migrate to $0$, resulting in link concentration and star network. Moreover, since non-native candidate delivers public work unconditionally, clients of $0$ get public work irrespective of who wins. In comparison, non-clients get public work only when the non-native candidate wins. Hence clients get more public work than non-clients.

It is interesting to look at comparative statics. Everything else remaining the same, if $b$ increases then we have fewer clients. Clientelism weakens because the inefficiency gap increases and hence the efficient non-native candidate becomes more attractive. However a decrease in $b$ beyond a limit also breaks clientism equilibrium. If $b$ is too low then it cannot be used as a clientelistic instrument. On the other hand, if the search cost interval contracts, everything else remaining the same, market transactions become more attractive and hence clientelism weakens.

The issue of commitment, elected elite delivering jobs as promised, has featured prominently in some theoretical models of clientelism (for instance Robinson and Verdier (2013)). Since our focus is on network structure, we simply assumed that contracts can not be reneged. However one can add a strategic move by elected elite in second period. Suppose she can allocate workfare job as she wishes at that point. However, it does not change our result because at that stage she is completely indifferent between choosing clients and non-clients. Our result characterizes equilibria where she chooses her clients.

\section{Empirical methods and further components of the data}

\subsection{Methodology}

Recall that our aim is to investigate the effects of interpersonal links, especially unidirectional links, and the role of clientelism on access to jobs under MGNREGS. The theoretical model predicted that concentration of unidirectional links and star-shaped network are some characteristics of clientelistic practices. Our model also predicts that clients get more public work jobs (MGNREGS) than non-clients. 

To explore these we estimated two sets of models. For the first set the outcome variable is the likelihood of getting jobs ever under the MGNREGS scheme and for the second set the outcome variable is the number of days worked (for each household) during the last 12 months under the MGNREGS scheme. Hereafter, the former will be called `participation' and the latter `days worked'. ‘Participation’ was measured by a question asking if anyone in the household had ever worked under MGNREGS scheme. If their answer was yes then this was coded as 1; 0 otherwise. Households were also asked the number of days they had worked (in total) under MGNREGS scheme over the last 12 months. Their answer was coded as ‘days worked’ variable. As this scheme is supposed to provide a maximum of 100 days of unskilled manual work to each rural household at a government stipulated minimum wage, any household reporting days worked greater than 100 was truncated to 100. 

We estimated models of the following form using OLS (or LPM in case of the binary outcome variables like ‘participation’), with (village) cluster robust standard errors and village fixed effects. In all models we included control variables (detailed in the next subsection) which we expected to influence access and demand for MGNREGS work. Therefore, the basic regression equation is of the following form:
\begin{equation}
y_{ij}=\alpha_0+\alpha_1 network-index_{ij}+\alpha_2 X_{ij}+\alpha_3 Z_j+\varepsilon_{ij}
\end{equation}
Where $y_{ij}$ is the MGNREGS employment variable for household $i$ in village $j$; $network-index_{ij}$ represents one of the network-related index for that household, $X_{ij}$ represents the household-level controls and $Z_j$ represents the village-level controls, the main variant being village fixed effect. 

First we estimated whether households which are receivers with unidirectional links, have better access to MGNREGS jobs than non-receivers of links and whether this is the case for receivers with reciprocal links as well. In this model we included the variable Linktype with non-receivers of links as the reference category (Model 1). We further investigate whether the intensity of these links matters. In Model 2 we included the degree variables (degree of reciprocal links and degree of unidirectional links) which, as we explained in Section 3.2, represent the number of households on which the sampled household has these types of links. In Models 3 and 4, we included the Concentration Index and the Weighted Concentration Index instead to get a better measure of intensity of these links. In these models we also control for degree of reciprocal links.

If we find evidence that unidirectional links matter, then it could be an indicator for underlying patron-client relations. To explore this we estimated whether being the client of a patron matters on the MGNREGS outcome variables (Model 5). Recall that a client is defined as a household which is recipient of at least one unidirectional link from a patron household and a patron household in a village is a provider of unidirectional links to at least 5\% of the sampled households in that village. As mentioned in Section 3.2, there is possibly measurement error in identifying patrons and thus, clients. In that case, some recipients of unidirectional links who should actually be clients could be mis-identified non-clients. So, in Model 6, we included a three category variable: non-recipient of unidirectional links, recipient of unidirectional links but not a client, client. 

We also wanted to know if there are certain types of patrons who were more effective in providing jobs for their clients. So, in Models 7 and 8 we test if it mattered whether the patron had political connections and whether the patron was in business. As MGNREGS is distributed by the local political administration, we expected a patron who is currently holding or has held a political position may be more influential in getting these jobs for their clients. So, in Model 7, we split the client group into those who are the client of at least one patron with some formal political position (current or in the past), and those who are the client of patrons none of whom have held such a position. For a similar reason, in Model 8, we test whether clients of at least one patron whose main occupation is some kind of business have better access to MGNREGS than clients of patrons with some other profession and non-clients.

In India, we know that caste plays a crucial role in social, economic and political spheres. In standard social group and social identity theory, members of the same social group behave more favourably towards each other than out-group members (Tajfel, 1989). So, we wanted to check whether it mattered whether the Pradhan/Sarpanch (the village administration senior leader) of the client’s village was of the same caste or a different caste. So, in Model 9 we split clients into 2 categories based on the Pradhan’s caste.  In all these models (Models 5-9) we also control for degree of reciprocal links.
	
\subsection{Control variables and regression samples}

In this section we discuss the control variables we included in our models which we expected to influence access and demand for MGNREGS work. We also specify the regression samples for the different models. 

The control variables--all at the household level--are:

Caste and religion: We have 5 categories: SC-ST-NT (Schedule Caste, Schedule Tribe, Nomadic Tribe), OBC (Other backward classes), Dominant Hindu castes (this includes Brahmin and other castes not included as SC, ST, NT or OBC), Muslim and uncodeable.

Presence of low-skilled workers: Number of household members between the age of 16-60 years who were educated up to secondary level at most.

Education: The highest educational level of the household. We have three categories: up to higher-secondary level (pre-university) education, under-graduation or equivalent degrees and above that.

Occupation: Whether the main occupation of a household is a stable one: coded as 1 if main occupation is running a business/factory/production unit, or salaried position in some organization, 0 otherwise. 

Remittance: Whether the household receives remittances from outside: 1 if it does and 0 otherwise.

Assets: (a) Amount of land owned in acres by the household, (b) Index of non-land assets owned by the household which ranges from 0 to 6--this is computed by aggregating 0-1 ownership indicators of 6 types of asset items - a `pucca' house of residence, additional house in a town or city, television, any automobile, expensive bed or palang and trees or plants that grow fruits or vegetables.

Community participation: (a) Whether the household has at least one member who either is (was) a member of local government or is a member of a political party/labour union; (b) whether the household mediates in community disputes; (c) whether the household has experience of visiting administrative officers or other such formal public institutions. All of these three are binary variables.

We restricted the analysis to households which were not patron households (there were 11 such households in our sample), and where the head of household was not a woman below the age of 18. For estimation of ‘participation’ models we included only those villages where there were at least 5 households that had ever carried out MGNREGS work. For estimation of ‘days worked’ models we included only those villages where at least 5 households had carried out MGNREGS work over the last 12 months. For these models we additionally restricted the sample to villages in Odisha and Maharashtra. This is owing to a issue of harmonization. Across the two fieldwork phases we made a change in the question wording: in the second phase instead of asking about the number of days worked under the MGNREGS scheme, we asked respondents to choose from 5 response categories (less than 25 days, 25-49 days, 50-74 days, 75-99 days, 100 or more days). The sample sizes are 2,770 and 1,267 households for the participation and days worked models.

We conducted some robustness checks as well which are discussed in Section 6.2

\subsection{Descriptive statistics}

In Table 3.3 we show means and proportions of the outcome and control variables for the two samples. The left panel shows the descriptive statistics for the ‘participation’ sample, and then split by client status. The right panel does the same for the ‘days worked’ sample. We find that a higher proportion clients (as compared to non-clients) report working under the MGNREGS scheme (65.3\% vs 40.8\%) and working for a higher number of days on average over the last 12 months (16.5 days vs 13.0 days). We also found that compared to non-clients client households were less likely to $(i)$ be of any dominant caste, $(ii)$ have any ‘stable occupation’ as the main occupation, $(iii)$ have at least one member who sends money home, $(iv)$ give advice to villagers or workers on own farm/business. Client households also have less wealth and assets on average. As these factors are likely to influence the demand for MGNREGS work, it is important to control for these variables in our models which we do. 

Table 3.4 shows the distribution of the different types of link variables and the different client status variables discussed in Section 5.1. As expected the degree of unidirectional links and concentration indices is higher for clients than non-clients. Among non-clients the higher proportion are non-receivers of links, followed by receivers of ‘unidirectional links’ and the lowest fraction consists of receivers of only ‘reciprocal links’. By definition clients are receivers of unidirectional links.

\section{Results}\label{result}

\subsection{The main findings}\label{main}

In Table 6.1 we report the estimated coefficients (and their standard errors) of the network-related variables estimated using OLS/LPM (with village fixed effects and village level cluster-robust standard errors) as specified in Models 1-4 (refer to Section 5.1). The left panel in the Table shows the results for ‘participation’ models and the right panel for the ‘days worked’ models. In Model 1 (for both participation and days worked) we find that receivers of unidirectional links have better access to MGNREGS work than receivers of reciprocal links and non-receivers of links. Specifically, compared to households which are non-receivers of links, households which are receivers of unidirectional links have a 13 percentage point higher chance of participating in MGNREGS and 3.33 more work days under MGNREGS programme in the last year. These are statistically significant at 5\% level and 10\% level respectively. The effects are substantial, around 28\% of the average probability of participation (0.46) in the ‘participation’ sample and 24\% of the average work days (14.1 days) in the ‘work days’ sample. The estimated coefficients for households with reciprocal links are 0 and 3.74 for participation and work days models but neither are statistically significant.

The estimated coefficients of degree of reciprocal and unidirectional links in Model 2 show a similar pattern with the coefficients for degree of unidirectional links being positive (0.09 and 1.76) and statistically significant at 5\% and 10\% levels of significance respectively. A similar pattern is observed for the two concentration indices (Models 3 and 4) with one exception: the coefficient for weighted concentration index in the days worked model is positive but not statistically significant.

This set of results confirms our expectation that unidirectional links and concentration of unidirectional links are crucial in getting access to MGNREGS jobs. Building on these results, we next explore the possible role of clientelism. As explained in Section 3.2, we used star-shaped networks to identify potential patrons and their clients. In Table 6.2 we provide the estimated coefficients of the different client status variables and their standard errors estimated using OLS/LPM (with village fixed effects and village level cluster-robust standard errors) for MGNREGS participation and days worked models (Models 5-9; refer to Section 5.1). Once again, the left panel shows the results for the ‘participation’ models and the right panel for the ‘days worked’ models.

First, as expected, we find that clients are more likely to participate ever in MGNREGS programme and such households work for more days under this programme. The coefficients translate into an increase in participation probability of 15\% of the sample average, and 31\% more number of work days than the sample average. These are statistically significant at 1\% level for participation and at 10\% level for work days. 

Next in Model 6 we test whether non-clients who report being recipients of unidirectional links also have better access to MGNREGS jobs as clients. We expect this to be the case as we may have failed to identify some patrons and hence their clients (See Section 5.1). We do find this to be case for the participation model but not for the days worked model (the coefficient was positive but not statistically significant even at 10\% level).

We then go on to test whether particular types of patrons are more influential in providing access of MGNREGS to their clients, specifically patrons with political connections (Model 7) and patrons whose main occupation is business (Model 8). We do find that patrons with political connections are able to provide their clients better access to MGNREGS than other types of patrons. But while business patrons are better able to provide MGNREGS work in terms of participation ever, business and non-business patrons are both able to provide more days worked in the last year. This finding is in apparent contrast to that by Maitra et al. (2020) who find differential impacts of political and business patrons in profitable use of micro-credit. But note that our outcome variable--workfare employment--is quite different in nature from which patrons with business are unlikely to generate any substantial additional profit.

Finally, we tested whether there was an in-group effect, that is, patrons of the same caste provided clients with better access than patrons of different caste (Model 9). We found this to be the case for days worked in the last year, but not for participation ever. So, there is some indication that both horizontal (along caste/religious lines) as well as vertical connections matter in allocation of such jobs.

\subsection{Robustness and sensitivity checks}\label{other}

We conducted a number of robustness and sensitivity checks. 

First of all, we re-estimated all models dropping village fixed effects and including some relevant village level characteristics instead. The findings are summarized in Tables A2, A3 in Appendix C.These village level control variables\footnote{For summary statistics see Table A1 in Appendix C.} are distance of the village from nearest town in KM, proportion of households whose main occupation is cultivation or agricultural labour, average rainfall in village in millimetre, fraction of net sown area irrigated and clientelism score of village. Recall from Section 3 that for each village the clientelism score is the weighted sum of unidirectional links provided by patrons to all their clients weighted by the number of their clients.  The pattern of the main results remains the same although the coefficients for receivers with ‘unidirectional links’ and degree of unidirectional links in days worked models (Model 1 and 2 in right panel of Table A2) are not statistically significant even at 10\% level of significance and degree of reciprocal links is significant at the 10\% level of significance for the participation model (Model 2 in left panel of Table A2).

Note that there might be two main factors, both of which are common in the literature, which might cast doubt on the direction of causality underlying our observed pattern of association. The first is reverse causality. In our context it might be argued that a household has unidirectional connections (or becomes a client) solely because {\em it has to seek MGNREGS employment, and thus has to establish connection with influential people in the locality.} The critique, therefore, is that demand for MGNREGS jobs produce unidirectional connection rather than the opposite. The second possible objection is that a household may create unidirectional connections as well as be inclined toward temporary workfare employment owing to some unobserved heterogeneity. Further, there could be routes different from what we have suggested for the transaction--quid pro quo through patron-client relation--for explaining the observed patterns of association.

For the second item of criticism mentioned above--that of unobserved factor(s)--we do not have any satisfactory response. 

But reverse causality is unlikely to be the factor driving our results. Recall that the MGNREGS scheme was introduced in 2006 and it became active in a full-fledged manner only in 2008 (i.e., about 4 years prior to our survey). Recall also from the introductory section that the average earning (in the last 12 months prior to survey) in our sample through the MGNREGS scheme is about INR 1500. The average incremental earning owing to unidirectional (or clientelistic) connection is about INR 400. While this sum is quite non-trivial for a poor household in rural India, the primary sources of income for such households (usually landless and with little assets) still remain, by and large, farming in rented land or working as casual labourers. Notice that in our survey we asked each household about one or two most important service providers in such (as well as in political and social) contexts. Therefore, a household approaching another `powerful' one only for such workfare jobs in relatively recent past and that particular connection proliferating into an important unidirectional relation in much more important dimensions of living does not seem to be quite realistic.

In support of our contention that indeed clientelistic provision of patronage is behind our results we performed the following robustness checks.          

While, as we indicated, MGNREGS jobs, our central outcome of interest, can act as a very appropriate tool for clientelistic control, we look for similar results for some other outcomes too. If our client variable does reflect a client-patron relationship then we should expect client status to have a negative effect on levels of unemployment (as they are more likely to get regular work via their patrons), and migration (as households where members have migrated to other places are less likely to be dependent on a patron-client relationship) for households whose main occupation is supplying casual labour. We measured level of unemployment by the number of months, as reported by the household-head in agricultural labour households, s/he had been unemployed in the year prior to the interview. We measured migration levels by the proportion of 18-60 year old household members living outside village. Further, a client household/a household with unidirectional connection is more likely to avail subsidized public health facilities rather than private medical centres/doctors (as without connection with powerful people, quality of service in such cheaper facilities in rural India is not assured at all). We would also expect clients to help out patrons during elections including distributing gifts in exchange for votes. Due to social desirability bias and security concerns respondents are less likely to answer questions about their involvement in such activities but are more likely to say yes to a proxy question about other people’s behaviour. So, we asked the question “Are there instances of gift giving before or during election in this village in the last 5 years?” to capture this behaviour. We expected clients to say yes to this question more than non-clients. We estimated these 4 models on both participation and days worked samples (See Table 6.3) and find results in the expected direction for vote buying, availing public medical facilities and number of months unemployed models (as well as for the migration model in the `ever worked' sample). This provides some partial but additional support to our claim that the variables we used to measure client do reflect a client-patron relationship.

It is possible that although patrons may be providing MGNREGS work to their clients selectively, but it is merely due to name-recognition rather than any underlying political-economic calculations. It has been observed--and recall that our conceptualization of clientelism is based on such observations--that clientelistic patronage tends to use perishable consumables like temporary jobs to retain patrons' control (Bardhan and Mookherjee (2012, 2017)) over the clients. In contrast, a one-time lump-sum favour is useless as a commitment device. If biased allocation of MGNREGS work is merely due to name recognition, then we should expect a similar pattern of association for one-time welfare benefits like BPL cards and assistance through Indira Awas Yojana (a one-time public grant for providing house to a poor household). We estimated the effect of being a client on the likelihood of being a recipient of such a one-time benefit by LPM with cluster robust standard errors. We included the same control variables and in one specification included village fixed effects and in another village level characteristics (See Table 6.4). We estimated the model on both the participation and days worked sub-samples. We do not find any client effect on these one-time transfers.

It is also conceivable that better awareness about the MGNREGS programme, rather than clientelism, is driving our results. Clients, being connected with patrons, perhaps, are better informed about MGNREGS work-schemes compared to non-clients. To check this possibility, we estimated the effect of being a client on an index of awareness about the MGNREGS by LPM with cluster robust standard errors. We included the same control variables and in one specification included village fixed effects and in another village level characteristics (See Table 6.4). We estimated the model on both the participation and days worked sub-samples. We do not find greater awareness about MGNREGS programme among clients than non-clients.

\section{Concluding Remarks}\label{conclusion}
We have used rich interpersonal interaction data, collected from a primary survey, to explore the link between patron-client relations and differential access to workfare jobs. We argue that our unidirectional network captures hierarchical power relations, as opposed to reciprocal networks of friendship/kinship relations. Moreover, patron-client relations give rise to the concentration of unidirectional links and star-like networks. We found that households with more unidirectional links or having a higher concentration of such links get more jobs. A household which receives some type of unidirectional link from the centre of a star-like network also gets greater access to jobs than others. Although our results do not nail down causal relationship, they remain robust under several sensitivity checks (including alternative specifications of client households,; which can be found in a working paper version of this work: Bhattacharya et. al.,  2017). Our paper uses social networks to identify the presence of rural clientelism and measures its impact on allocation of scarce resources. The twin aspect of the present exercise is that the presence or intensity of such institutions should inhibit the functioning of those services and items of infrastructure which might reduce the power of the elites in controlling their clients. For example, the spread of formal banking and credit institutions would weaken the role of informal rural credit, the increase of public irrigation would weaken the necessity of getting access to irrigation-water through private means, etc. In a companion paper (Bhattacharya, Kar and Nandi, 2021), based on the same data used in this work, we have found supporting evidence of this feature.

\pagebreak
\pagestyle{plain}
\hypertarget {appendix A}
\appendix

\Large{

\noi{\bf{Appendices}}

}

\normalsize

\section{Proof of Theoretical Results}
{\bf Proof of Proposition 1}\\
We compute subgame perfect Nash equilibrium by backward induction. During election, a poor voter chooses between candidates. Since voting uses secret ballot, an unlinked agent has no incentive to vote for an elite. His support will go unnoticed and the elected elite will only reward her clients. Thus it is the dominant strategy for unlinked poor to vote for the non-native candidate, $N$. Similarly a linked poor never votes for an elite other than his patron. He can however vote for the non-native candidate. \\
Suppose $k$ is a client of elite $0$. If $k$ votes for the non-native candidate instead of his patron then winning probability of the non-native candidate increases by $\frac{1}{n}$ and that of elite $0$ decreases by the same margin. When the non-native candidate wins, she allocates maximum amount of public work to everyone, including $k$. Thus $k$'s payoff increases by $\frac{1}{n}b$. On the other hand $k$'s contract with $0$ generates the following payoff: if $0$ wins then $k$ gets $\theta b$ amount of public work, otherwise $k$'s link is severed and $k$ has to access resources through market in the second period. Thus if $0$ is not elected then $k$'s payoff decreases by $(\theta b+2s_k)$. Therefore $k$ votes for his patron $0$ if and only if $$\left [\frac{1}{n}b-\frac{1}{n}(\theta b+2s_k)\right ]\leq 0 \Leftrightarrow s_k\geq \frac{b(1-\theta)}{2}$$
By exactly same arguments, $1$'s clients (or $2$'s clients) vote for their patron if and only if $s_k\geq b(1-\theta)$. A crucial difference is that $1$ (resp. $2$) offers only one resource. Thus when $1$ (resp. $2$) is not elected payoff of his client only decreases by $(\theta b+s_k)$.

From the objective function of elites, we know that an elite always wants an extra client if and only if the client votes for her. Thus in the first period, $0$ gives consent to all agents whose search cost is above $\frac{b(1-\theta)}{2}$, that is agents in $\Pi_0$. Similarly $1$ and $2$ give consent to all agents in $\Pi_1$ for link formation. Note that $\Pi_0\supset \Pi_1$.  Agents in $\Pi_U$ do not have any offer, so they remain unlinked and vote for the non-native candidate. From now on we assume that $b(1-\theta)<2$, otherwise we are done. To complete part $(a)$, we must show that all agents in $P_0$ accept $0$'s offer, that is, establish link with $0$. This is done in two steps.\\
Step 1: We show that all agents in $\Pi_0$ prefer to link with $0$ than to remain unlinked. Since agents in $\Pi_0\setminus \Pi_1$ only have $0$'s offer, it means that they accept it.\\
Step 2: We show that each agent in $\Pi_1$ prefers to link with $0$ than with $1$ and $2$ because he believes that the rest of $\Pi_0$ are going to connect to $0$.

Proof of Step 1: This part itself has two steps. Suppose $j$ has the highest search cost among all poor who are not in $\Pi_1$. That is $s_j=\max_k\{s_k\mid s_k< \min\{1,b(1-\theta)\}$. Since search costs appear in a grid of $\frac{1}{n}$, we know $[\min\{1,b(1-\theta)\}-s_j]\leq \frac{1}{n}$. Size of $\Pi_0\setminus \Pi_1$ is $\left [\min\{1,b(1-\theta)\}-\frac{b(1-\theta)}{2}\right ]$. By $(a3)$, $[\min\{1,b(1-\theta)\}-s_j]\leq \frac{1}{n}\leq \left [\min\{1,b(1-\theta)\}-\frac{b(1-\theta)}{2}\right ]$ and hence $j\in \Pi_0$. Thus $j\in \Pi_0\setminus \Pi_1$\\
We show that $j$ prefers to link with $0$ than to remain unlinked, irrespective of what others choose. If $j$ links with $0$ in period 1, we know that he will later vote for $0$ because $j \in \Pi_0$. $j$'s lifetime expected payoff from linking with $0$ is $\left [\alpha(\theta b)+(1-\alpha)(-2s_{j})+\beta b-c\right ]$, where $\alpha$ is the win probability of $0$ and $\beta$ is the win probability of $N$. Win probability of other candidates do not appear in this expression, because they won't allocate public work to $j$. Agent $j$ also pays the link cost. Instead if he declines the offer and remains unlinked then later he votes for $N$. Agent $j$'s lifetime expected payoff from that option is $\left[-4s_{j}+\left ( \beta+\frac{1}{n}\right )b\right ]$ - he pays search costs in both periods and $N$'s win probability increases by $\frac{1}{n}$.
The payoff difference is,
\begin{equation}\label{app1}
\Delta=\bigg (\alpha(\theta b)+(1-\alpha)(-2s_{j})+\beta b-c\bigg )-\left (-4s_{j}+\left ( \beta+\frac{1}{n}\right )b\right)
\end{equation}
Since the above expression is increasing in $\alpha$, it will be enough to show that $\Delta>0$ at $\alpha=\frac{1}{n}$. This will complete the proof that $j$ prefers to link with $0$ than to remain unlinked, irrespective of what others choose.
\begin{align*}
\Delta &=\left (\frac{1}{n}(\theta b)+\left (1-\frac{1}{n}\right )(-2s_{j})+\beta b-c\right )-\left (-4s_{j}+\left ( \beta+\frac{1}{n}\right )b\right)\\
&=2\left (s_j-\frac{c}{2}\right )+\frac{2}{n}\left (s_j-\frac{b(1-\theta)}{2}\right )\\
&=2\left [\min\{b(1-\theta),1\}-\frac{1}{n}-\frac{c}{2}\right ]+\frac{2}{n}\left [s_j-\frac{b(1-\theta)}{2}\right ] \geq 0
\end{align*}
The last inequality follows from $(a3)$, which gives $\left[\min\{b(1-\theta),1\}-\frac{1}{n}-\frac{c}{2}\right ]\geq 0$ and from the fact that $j\in P_0$, which implies $\left [s_j-\frac{b(1-\theta)}{2}\right ]\geq 0$.

Now consider $i$ whose search cost $s_i=\left (s_{j}-\frac{t}{n}\right )$ and $i\in \Pi_0$. That is $i$ comes $t$ place before $j$. We shall show that $i$ prefers to link with $0$ than to remain unlinked. We prove this by induction on $t$. We already know that the statement is true for $t=0$. Assume that it is true up to $(t-1)$. Thus $j,(j-1),\ldots, (j-(t-1))$ prefer to link to $0$ than to remain unlinked. We shall now prove the same for $(j-t)$. Agent $(j-t)$'s expected payoff difference between accepting $0$'s offer and remaining single is similar to Equation $\ref{app1}$.
\begin{align*}
\Delta &=\left [\bigg (\alpha(\theta b)+(1-\alpha)(-2s_{j-t})+\beta b-c\bigg )-\left (-4s_{j-t}+\left ( \beta+\frac{1}{n}\right )b\right)\right ]\\
&=\left [\alpha\left (\theta b+2s_{j-t}\right )+2s_{j-t}-\left (\frac{1}{n}b+c\right )\right ]=\left [\alpha\left (\theta b+2s_{j}-\frac{t}{n}\right )+2\left (s_{j}-\frac{t}{n}\right )-\left (\frac{1}{n}b+c\right )\right ]\\
&=\left [\left (\alpha-\frac{1}{n}\right )\left (\theta b+2s_{j}-\frac{t}{n}\right )-\left( \frac{t}{n^2}+\frac{2t}{n}\right )\right ]+\left [\frac{1}{n}\left (\theta b+2s_{j}\right )+2s_{j}-\left (\frac{1}{n}b+c\right )\right ]
\end{align*}
The second term in the above expression is exactly $j$'s payoff difference between linking with $0$ unilaterally and remaining unlinked. We have already shown that it is positive. Since, $s_{j}>\frac{b(1-\theta)}{2}$, the first term
\begin{align*}
\left [\left (\alpha-\frac{1}{n}\right )\left (\theta b+2s_{j}-\frac{t}{n}\right )-\left( \frac{t}{n^2}+\frac{2t}{n}\right )\right ]&\geq \left [\left (\alpha-\frac{1}{n}\right )\left (\theta b+b(1-\theta)-\frac{t}{n}\right )-\left( \frac{t}{n^2}+\frac{2t}{n}\right )\right ]\\
&=\left [\left (\alpha-\frac{1}{n}\right )\left (b-\frac{t}{n}\right )-\frac{t}{n}\left( \frac{1}{n}+2\right )\right ]\\
&\geq \frac{t}{n}\left[ b-\left (\frac{t+1}{n}+2\right )\right ]> 0
\end{align*}
Note that $\alpha$ is $0$'s win probability including $(j-t)$'s vote. From induction step we know $t$ other individuals, $j,(j-1),\ldots, (j-(t-1))$ vote for $0$. Thus $\alpha\geq \frac{t}{n}+\frac{1}{n}$. The inequality $\left[ b-\left (\frac{t+1}{n}+2\right )\right ]> 0$ follows from $b\geq 3$ that is restriction $(a1)$. This completes Step 1.

Proof of Step 2: Next we show that agents in $\Pi_1$ prefer to connect to $0$ rather than link to $1$ or $2$. If $b(1-\theta)\geq 1$ then $\Pi_1$ is anyway empty and we have proved our result. We assume $b(1-\theta)<1$.\\
In step 1, we have proved that $\Pi_0\setminus \Pi_1$, which is of size $\frac{b(1-\theta)}{2}$, link with and vote in favour of $0$. Since $b(1-\theta)<1$, $\Pi_1$ is also non-empty. Thus in the proposed equilibrium, probability that $0$ wins, denoted by $\alpha_0$, is at least $\left [\frac{b(1-\theta)}{2}+\frac{1}{n}\right ]$. Let us first show that there is no unilateral deviation from the proposed equilibrium by agents in $\Pi_1$. If $k\in \Pi_1$ links with $0$, then his lifetime payoff is $\left [\alpha_0(\theta b)+\left (1-\alpha_0\right )(-2s_k)+\beta b-c\right ]$, where $\beta$ is win probability of $N$. If $k$ unilaterally deviates and link with $1$ (or $2$), then his payoff is $\left [\frac{1}{n}\left (\theta b \right )+\left (1-\frac{1}{n}\right ) \left (-s_k\right )-\beta b - 2s_k\right ]$. This is similar to the last expression, except two elements - punishment cost is now $s_k$ and in both the periods $k$ has to buy one resource from the market, costing $2s_k$. The payoff difference can be rewritten as
\begin{align}\label{app2}
\Delta &=\left (\alpha_0-\frac{1}{n}\right )\left (\theta b+2s_k\right )+\frac{1}{n}s_k - \left (c-s_k\right )
\end{align}
$\Delta$ is increasing in $s_k$. If we can show that $\Delta$ is positive for the lowest $s_k$ in $\Pi_1$, then we are done. Computing $\Delta$ at $s_k=b(1-\theta)$, we get
\begin{align*}
\Delta &=\left (\alpha_0-\frac{1}{n}\right )\big (b+b(1-\theta)\big)+\frac{1}{n}b(1-\theta) - \big(c-b(1-\theta)\big)\\
&\geq 3\left [\frac{b(1-\theta)}{2}\right ]-b(1-\theta)\geq 0
\end{align*}
The second last inequality follows from the following observations - $(i)$ $\left (\alpha_0-\frac{1}{n}\right )\geq \frac{b(1-\theta)}{2}$, $(ii)$ $b>3$ by $(a1)$ and $(iii)$ $[c-b(1-\theta)]<b(1-\theta)$ by $(a2)$.\\
Therefore the partition of $\left [\Pi_0\cup \Pi_U\right ]$ is a subgame perfect Nash equilibrium. All agents in $\Pi_0$ are linked to $0$ and vote for $0$, while the rest are unlinked and vote for $N$. Clients of $0$ receive public work $\theta b$ when $0$ wins and receive $b$ when the non-native candidate wins. Whereas agents in $\Pi_U$ do not receive any public work when $0$ wins. Thus clients get more public work than non-clients.

Finally, we find a sufficient condition under which the above is the only equilibrium. Suppose $\alpha_0$ and $\alpha_1$ denote the win probability of elite $0$ and $1$ respectively in equilibrium, We already know that $\Pi_0\setminus \Pi_1$ become $0$'s client and $\Pi_U$ remain non-client. Thus $\alpha_0\geq \frac{b(1-\theta)}{2}$ and $\alpha_1\leq \left [1-(b(1-\theta)\right ]$. We show that an agent in $\Pi_1$ prefers to link to $0$ when $b(1-\theta)\geq \frac{6}{7}$.  Let us compute how much $k\in \Pi_1$ gains when he chooses $0$ instead of $1$. As in Equation $\ref{app2}$,
\begin{align*}
\Delta &=\left (\alpha_0+\frac{1}{n} -\alpha_1\right )\big (\theta b+2s_k\big )+\alpha_1s_k - \big(c-s_k\big )
\end{align*}
Once again this expression is increasing in $s_k$. Thus we show that $\Delta$ is positive at the lowest possible $s_k$, which is $b(1-\theta)$.
\begin{align*}
\Delta &=\left (\alpha_0+\frac{1}{n}-\alpha_1\right )\big (b+b(1-\theta)\big )+\alpha_1\big (b(1-\theta)\big ) - \big (c-b(1-\theta)\big )\\
&\geq 3\left [\frac{b(1-\theta)}{2}-\big(1-b(1-\theta)\big )\right ]-b (1-\theta)\\
&=\left [\frac{7b(1-\theta)}{2}-3\right ]\geq 0
\end{align*}
This complete the proof of Proposition 1.

\newpage

\section {Sample Design}

Our survey (``Local Institutions and Rural Economic Performances" (LIREP)) sample has a multi-stage, clustered and stratified design. The target sample size was 3600 households based on cost restrictions. As mentioned above, one of the key information that this survey aimed to collect was the local dependence structure and so it was essential to collect information from all or a sizeable number of households in each village. So,  As a result it was decided to select and interview approximately 100 households from each of the selected villages which meant that 36 villages could be selected in the sample.

India is a vast country with 29 states and 7 union territories and each of these regions are culturally and politically different with many policies being implemented at the regional level. To be able to control for these state level effects it was decided to confine the sample to three states (with 12 villages from each state) so that we had sufficient sample sizes at the state level. The three states chosen were Odisha, Maharashtra and (the Eastern part) of Uttar Pradesh (UP). These three states or sub-state regions were chosen for the following reasons. First, these states are located in three different parts of India: Maharashtra in the west, Odisha in the east and UP in the northern part. Secondly, each of these is a major state in India with each having a major different predominant language-based ethnic group. Third, these states have experienced several different historical types of administrative and land-revenue systems during the colonial period (permanent settlement, princely states, taluqdari systems, ryotwari system) which have been shown to have affected the development of post-colonial institutions, and in turn, current economic outcomes. Finally, measuring the institutional impacts of left-wing extremist (LWE hereafter) politics on rural household outcomes was one of our research goals. So, another reason for choosing these states is that LWE activities are prevalent in some parts of each of these states.\\

\noindent {\em Stage 1: Selecting blocks using a stratified design}\\

\noindent Our main research goal was to measure the effect of local level dependence structure on rural household and village level outcomes. But as these local level dependence structures are known to vary by a number of regional factors, we decided to increase the variability of the sample in terms of the local level dependence structures by stratifying the sample by these factors. To increase the variability of the sample along a number of characteristics and to ensure enough sample sizes for one of the variables of interest, left wing extremism, it was decided to stratify the sample along the characteristics as specified below. Most of the information about these stratification variables were available either at the district or at the block (a smaller geographical unit than the district) level. So, it was decided to first select blocks from each of the different strata using probability proportion to size (PPS) sampling where size was measured by the number of households in the block (as in 2001 Census of India, the latest that was available to us) and then select a village randomly from the selected blocks again using PPS sampling method where size was measured by the number of households in the village. The characteristics we used for stratification for each state subsample were as follows:
\begin{itemize}
\item Whether the block had experienced left wing extremist activities (L) or not (NL) between the period 2005 to 2010. This was identified using a number of different sources. 
\item Whether the district containing the block was in a coastal (C) or non-coastal region (NC). This was done directly by using maps. Coastal regions were are expected to have occupational diversity while people in more interior regions (non-coastal) are expected to be mainly in agricultural occupation. To be able to identify different types of dependence, not only predominantly agriculture-based dependence links, the sample was also stratified by coastal and non-coastal region.
\item Whether historically the district was under ryotwari (R) or non-ryotwari (NR) system during the colonial rule. This was identified using the classification provided by Banerjee and Iyer (2005). These historical administrative and land-revenue systems are known to affect post-colonial institutions.
\end{itemize}

Every state did not include all 8 combinations, so we ended up with 13 mutually exclusive and exhaustive strata within the three states (see the Table below). with the added constraint that 12 blocks would have to be selected from each region. As some analysis would look at the LWE impact we also decided that there should be a sufficient number of villages from the LWE stratum. We also needed to select 12 blocks from each state. Our resulting stratification strategy is described in the Table below.

\begin{tabular}{cccc}

\hline 

Stratum No.&State&Stratum&Number of blocks\\

\hline

1&	Eastern UP&	L,NC,NR&	4\\
2&	Eastern UP&	NL,NC,NR&	8\\
3&	Odisha&	L,CO,NR&	2\\
4&	Odisha&	L,NC,NR&	3\\
5&	Odisha&	L,NC,RY&	1\\
6&	Odisha&	NL,CO,NR&	2\\
7&	Odisha&	NL,NC,NR&	3\\
8&	Odisha&	NL,NC,RY&	1\\
9&	Maharashtra&	L,NC,NR&	4\\
10&	Maharashtra&	NL,CO,NR&	1\\
11&	Maharashtra&	NL,CO,RY&	2\\
12&	Maharashtra&	NL,NC,NR&	2\\
13&	Maharashtra&	NL,NC,RY&	3

\end{tabular}

\vskip2em

\noindent {\em Stage 2: Assigning selected blocks to forest and non-forest sub-samples}\\

\noindent At the next sampling stage was we selected one village from each selected block. In the first sampling stage one of the variables we had stratified by was LWE activity. But as blocks are large areas with on average 170 villages (and 50\% of blocks have more than 150 villages but 99\% of blocks have less than 550 villages), not all villages in an LWE affected block will be affected by LWE activity. As it was extremely difficult to get precise information on exactly which of the several hundreds of villages in a block has a history of LWE activities, we decided to indirectly screen for LWE affected villages by selecting villages in these LWE affected blocks that were very near to forest. We used this strategy because forest cover has been found to be highly correlated with LWE activity at least at the district level and there is anecdotal evidence that LWE organisations mainly base their activities in dense forests as state forces find it difficult to enter these areas. We decided to draw two sub-samples from these LWE affected blocks - one from areas next to forests and the other from areas away from forests. We did this by collecting maps of forest cover from the Geological Survey of India and the Forest Research Institute and then overlaying thoese on the maps of villages. We decided to assign the following number of blocks to the forest and non-forest sub-samples of the LWE based strata.

\begin{itemize}

\item Eastern UP- L,NC,NR stratum:: As one of the 4 selected blocks in this strata no forested village, this block was automatically assigned to the Non-forested Sub-Sample and the remaining blocks in that strata, since they summed up to the assigned number of blocks for the Forest Sub-Sample, were allocated to the Forest Sub-Sample.
\item Odisha - L,CO,NR stratum: As one of the selected blocks had no forested village, this block was automatically assigned to the Non-forested Sub-Sample and the remaining blocks in that strata, since they summed up to the assigned number of blocks for the Forest Sub-Sample, were allocated to the Forest Sub-Sample. 
\item Odisha - L,NC,NR stratum: We selected 2 out of the 3 blocks by PPS, where size measure was the proportion of households in forested villages in these blocks, for the Forest Sub-Sample.
\item Odisha - L,NC,RY stratum: The only selected block from this stratum was automatically assigned to the Forest Sub-Sample.
\item Maharashtra - L,NC,NR strata: We selected 3 out of the 4 blocks by PPS where size measure was the proportion of households in forested villages in these blocks, for the Forest Sub-Sample.

\end{itemize}

\noindent {\em Stage 3: Selecting villages from selected blocks}\\

\noindent Finally we selected one village from each of the 36 selected blocks using PPS where size is measured by the total number of households in the village. In two of the villages selected in Maharashtra, battle between the Indian Security Forces and the LWE militants raged  when the HH household survey was to start (in March-April, 2013). So, we had to replace these two villages by two other (safer) villages in the same or adjacent districts with similar administrative and land-revenue history and similar geographical features.\\ 

\noindent {\em Stage 4: Selecting households from selected villages}\\ 

\noindent In villages where the total number of households was less than 100, all households were selected for survey. In each village where the total number of households was more than 100, upto 110 households were selected using simple random sampling. The sampling frame used was the most recent electoral roll for those villages. The target was to interview at least 100 households in each village and at most 110 households. In cases, where due to attrition, non-response etc with the initially chosen sample of HHswe could not reach the target sample size of 100, additional households were selected from the remaining households in the village again using simple random sampling to reach the target sample size. This was repeated until the target sample size was reached. In the final sample, 21 of the sampled villages included less than 50\% of the HHs in the villages, 5 included 50-60\% of the HHs in the villages, 3 included 60-70\% of the HHs in the villages, 2 included 80-95\% of the HHs in the villages and 4 were village censuses.

\newpage

\begin{figure}[htbp]

\begin{center}

\includegraphics[angle=0, scale=0.7]{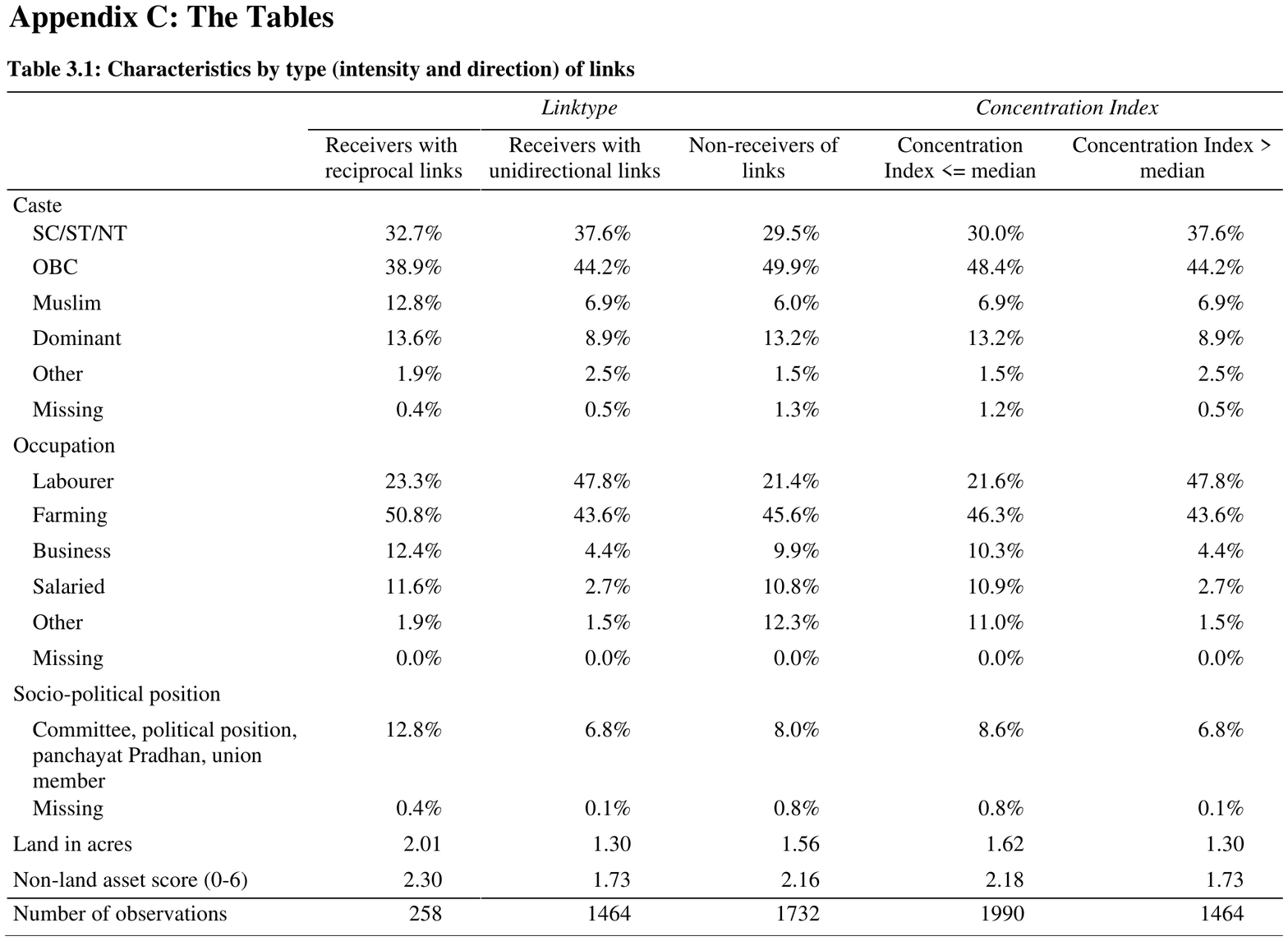}
\label{default}

\end{center}

\end{figure}

\newpage

\begin{figure}[htbp]

\includegraphics[angle=0, scale=0.7]{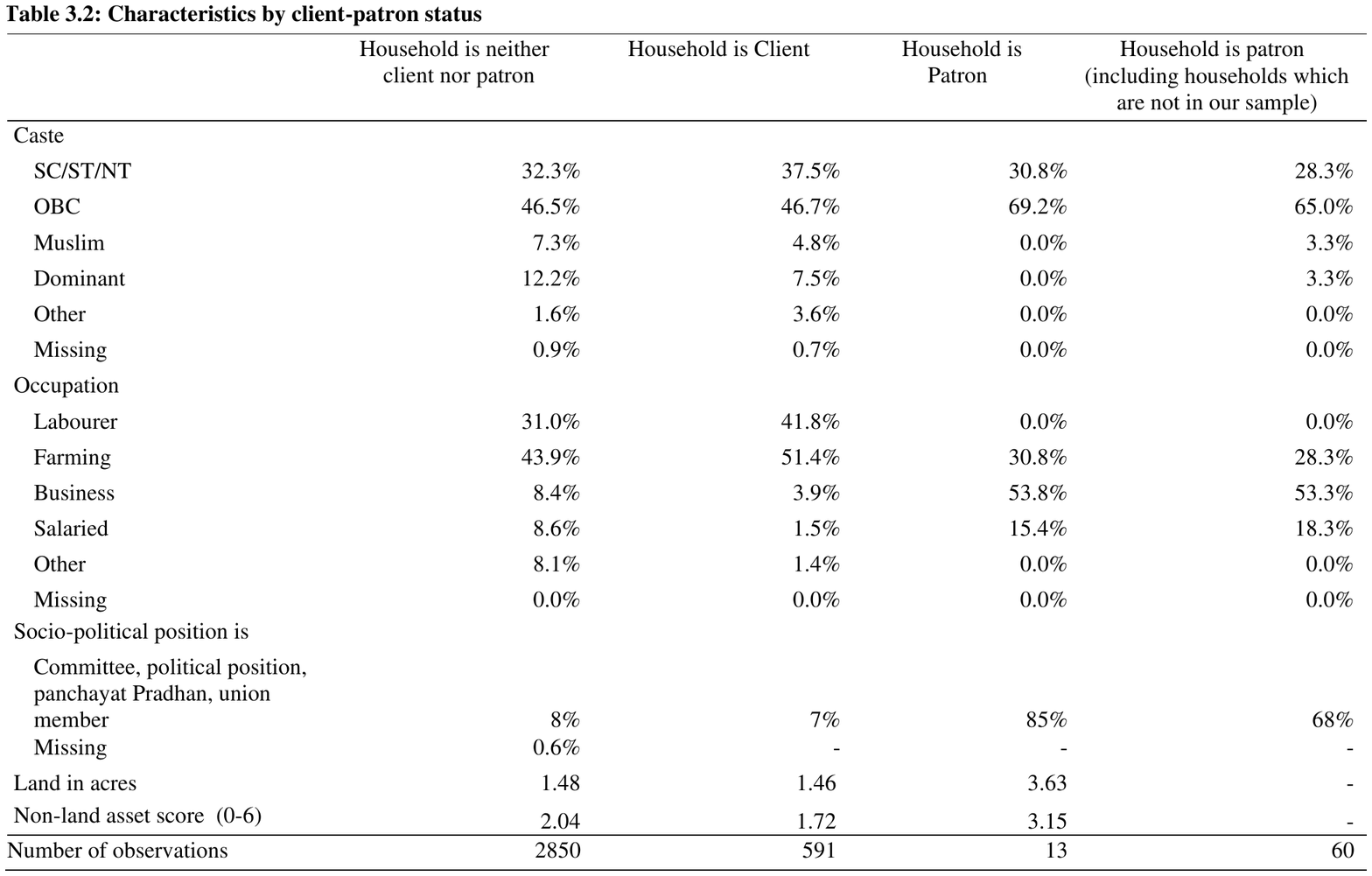}
\label{default}

\end{figure}

\newpage

\begin{figure}[htbp]

\includegraphics[angle=90, scale=0.7]{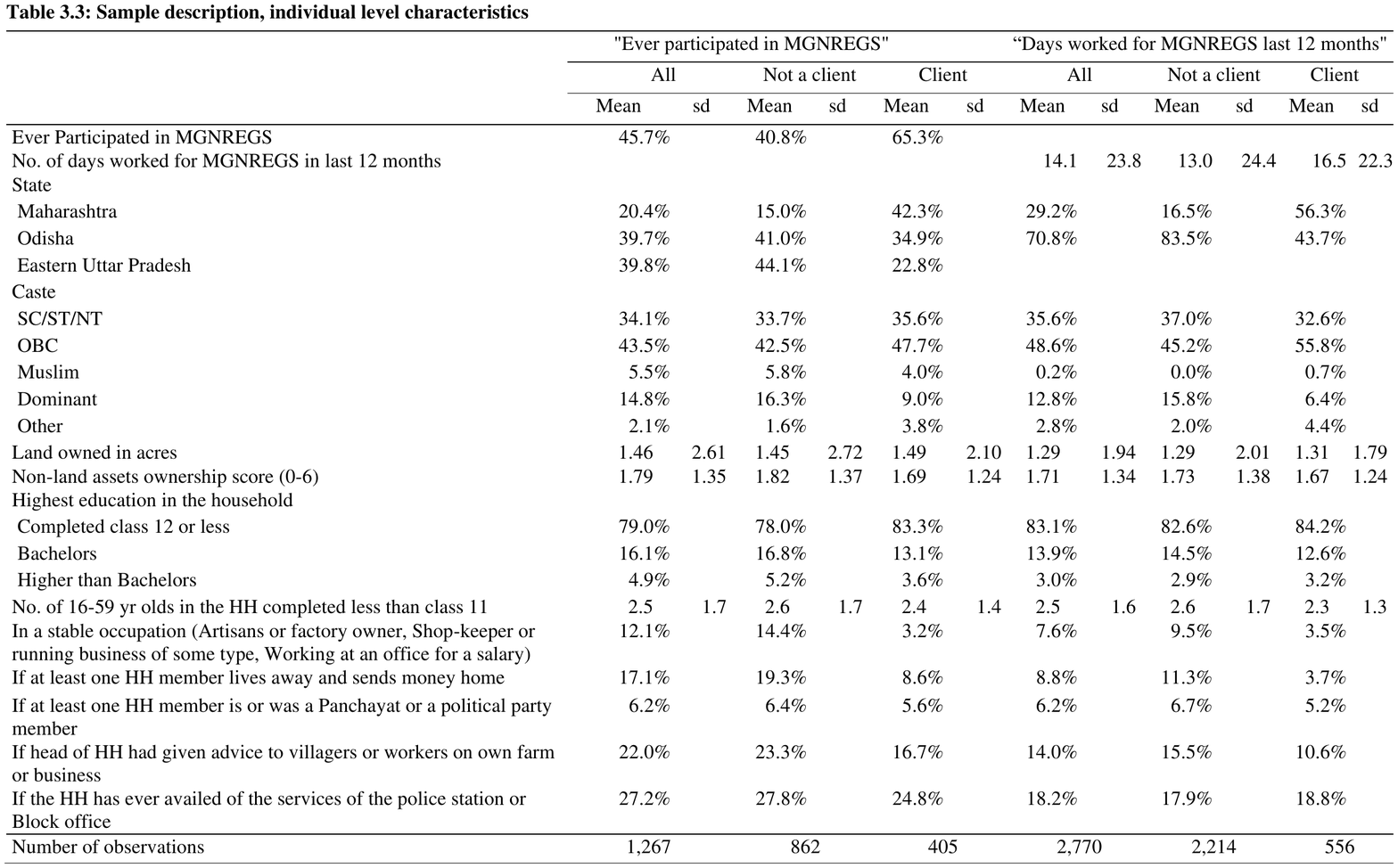}
\label{default}

\end{figure}

\newpage

\begin{figure}[htbp]

\includegraphics[angle=90, scale=0.7]{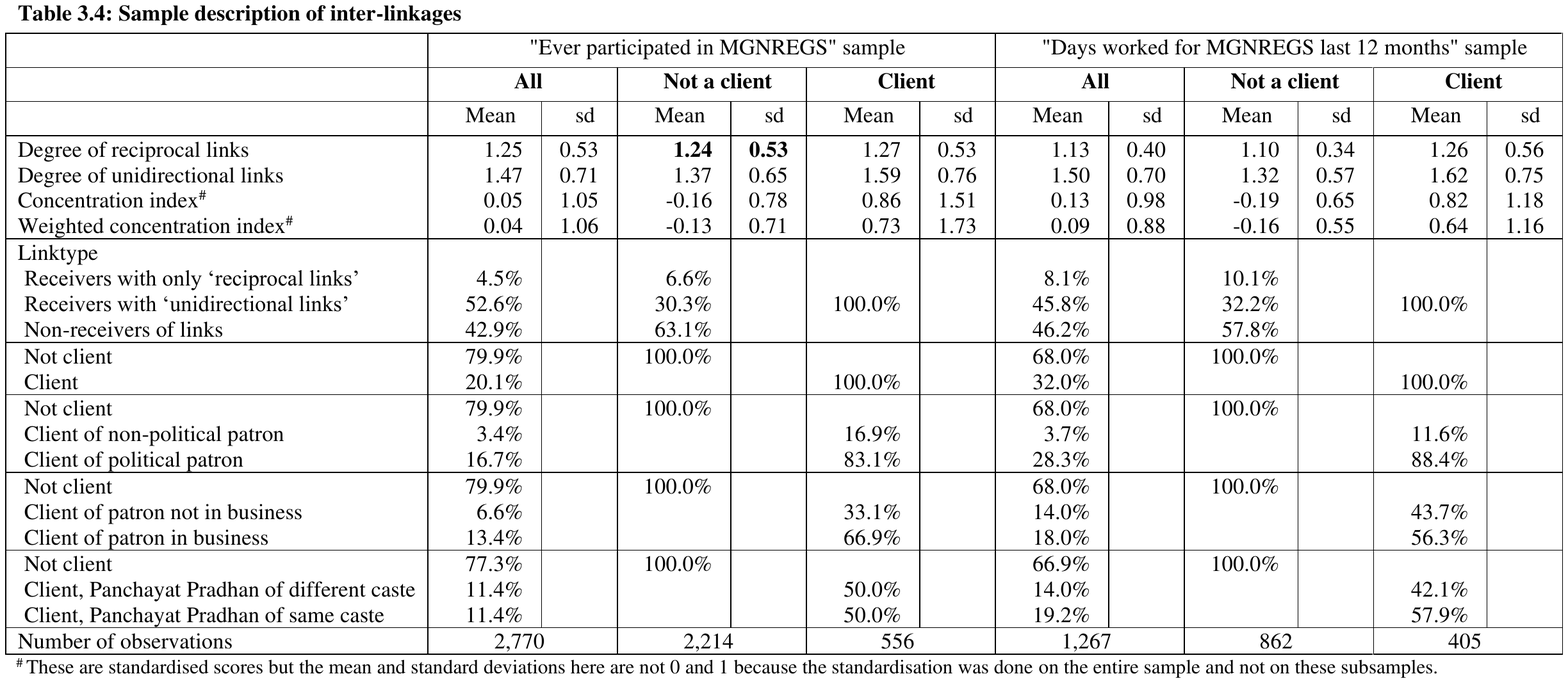}
\label{default}

\end{figure}

\newpage

\begin{figure}[htbp]

\includegraphics[angle=90, scale=0.7]{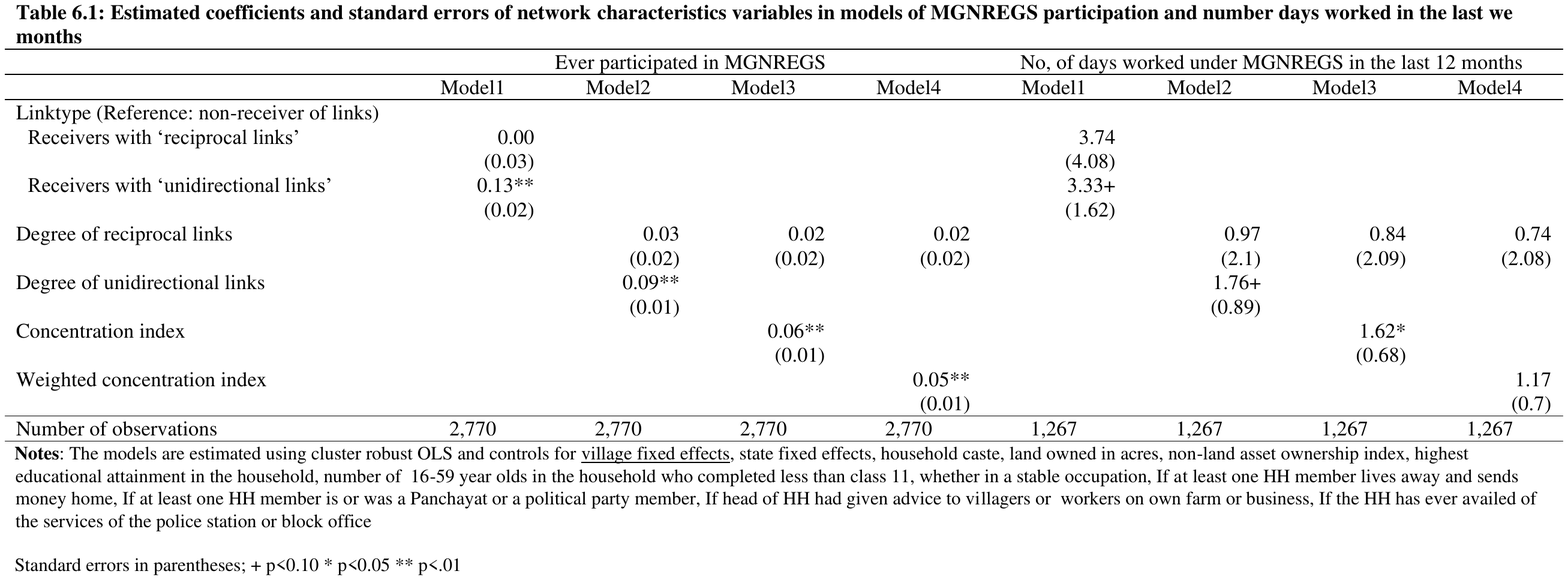}
\label{default}

\end{figure}

\newpage

\begin{figure}[htbp]

\includegraphics[angle=90, scale=0.7]{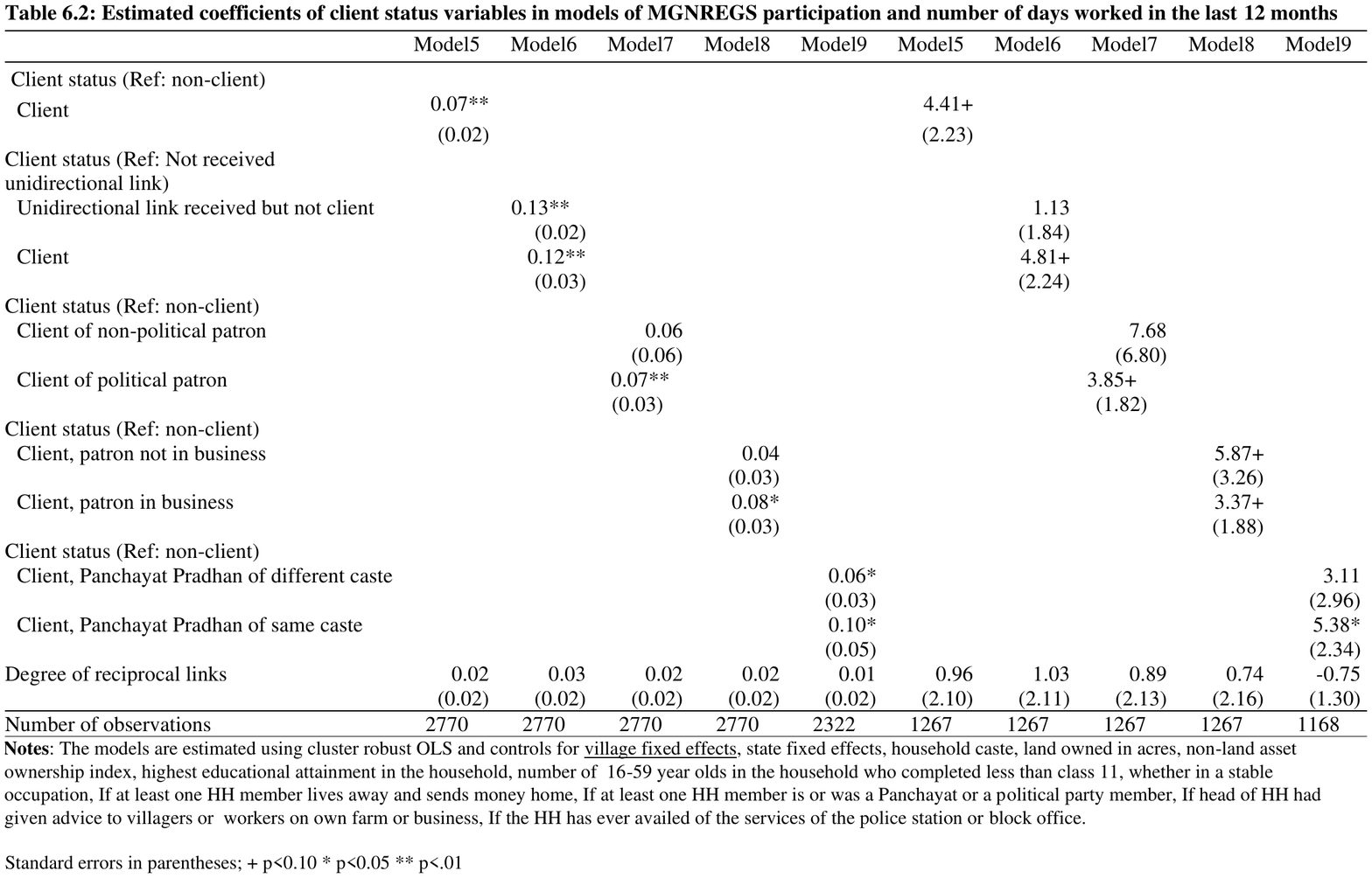}
\label{default}

\end{figure}

\newpage

\begin{figure}[htbp]

\includegraphics[angle=90, scale=0.7]{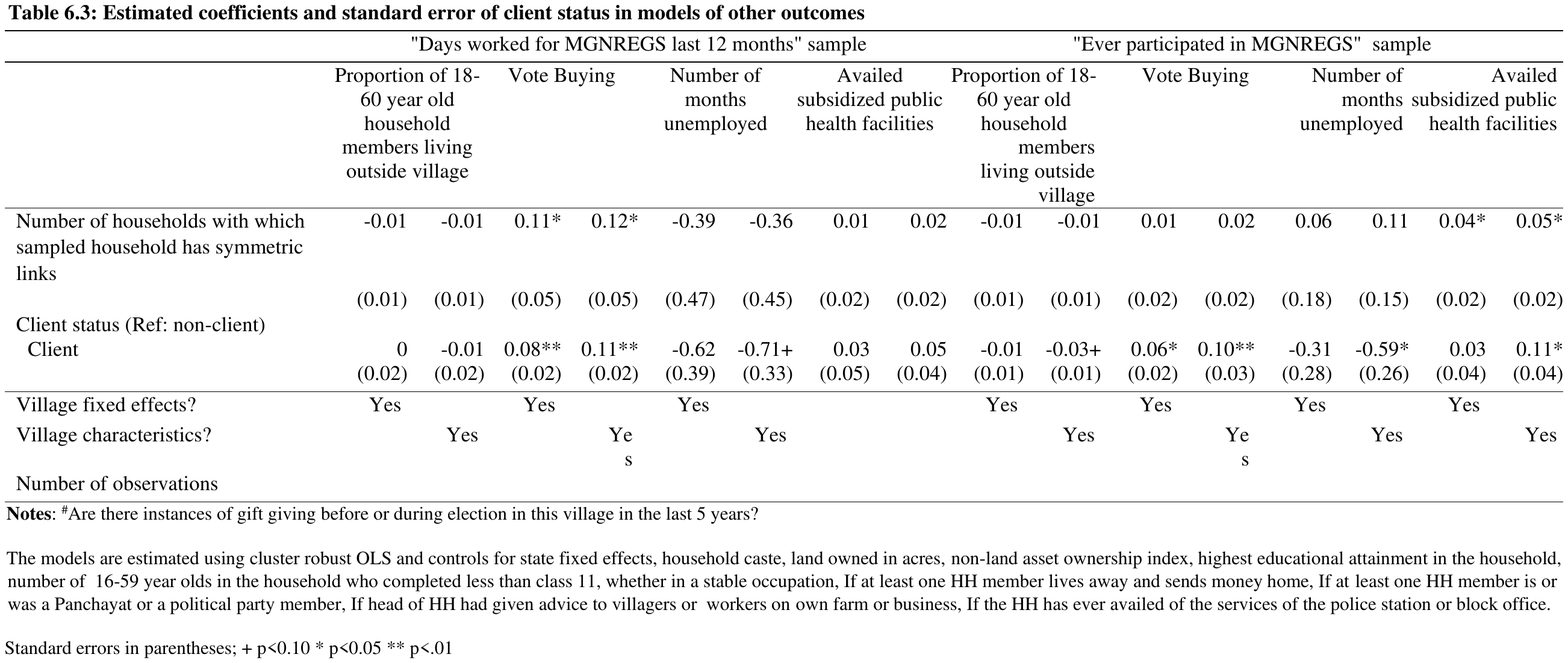}
\label{default}

\end{figure}

\newpage

\begin{figure}[htbp]

\includegraphics[angle=90, scale=0.7]{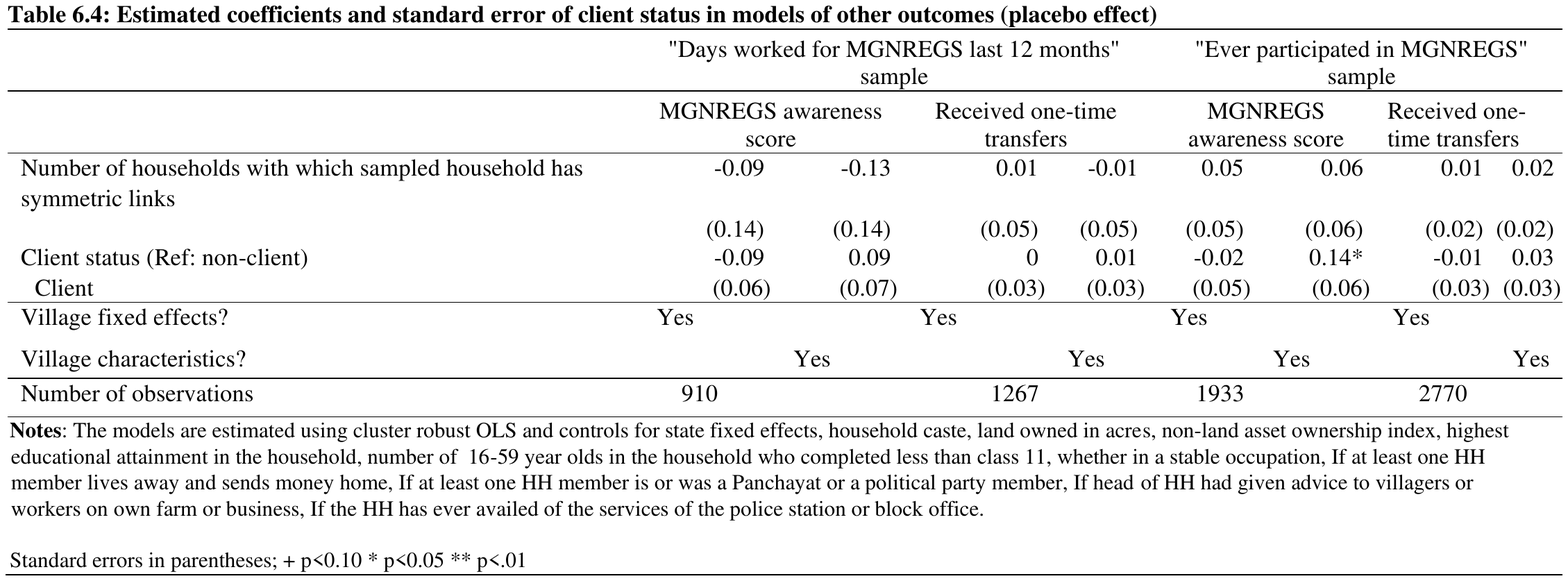}
\label{default}

\end{figure}

\newpage

\begin{figure}[htbp]

\includegraphics[angle=0, scale=0.7]{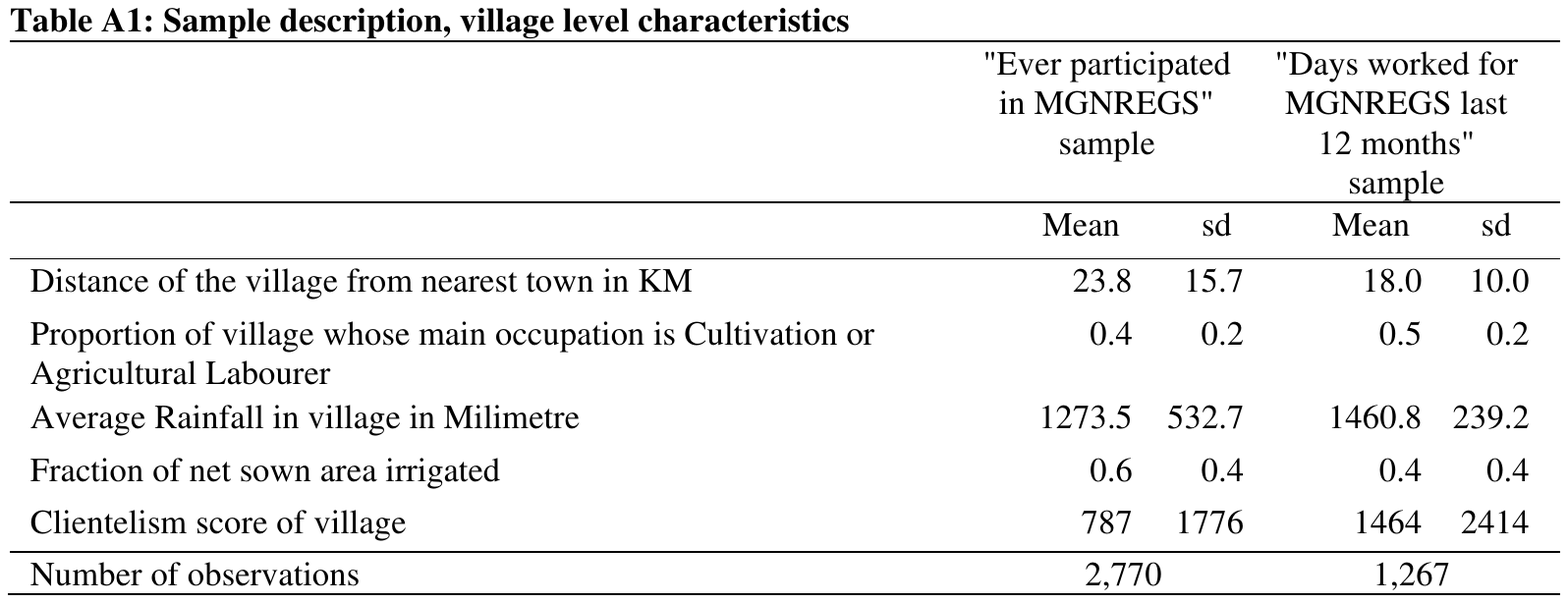}
\label{default}

\end{figure}

\newpage

\begin{figure}[htbp]

\includegraphics[angle=90, scale=0.7]{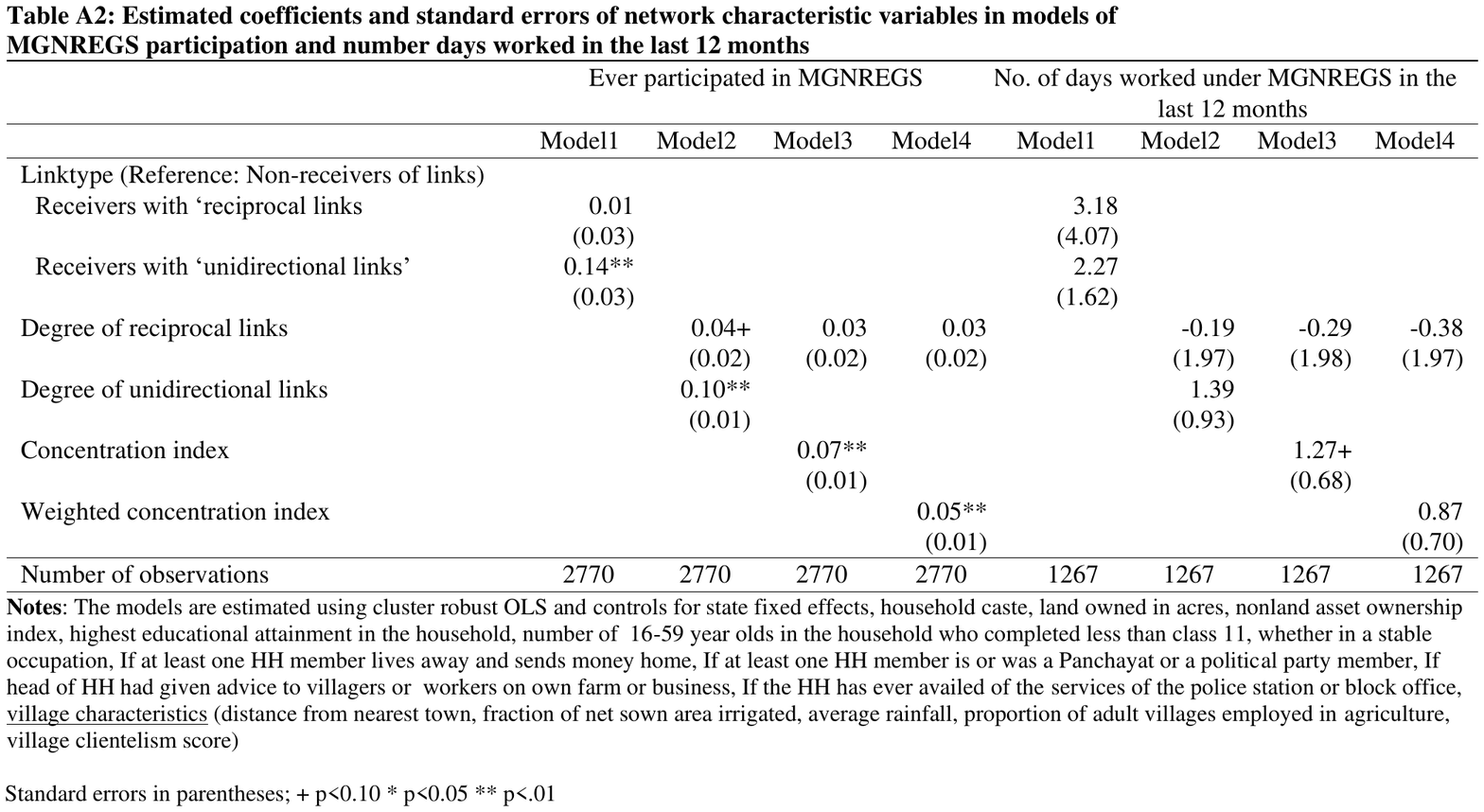}
\label{default}

\end{figure}

\newpage

\begin{figure}[htbp]

\includegraphics[angle=90, scale=0.7]{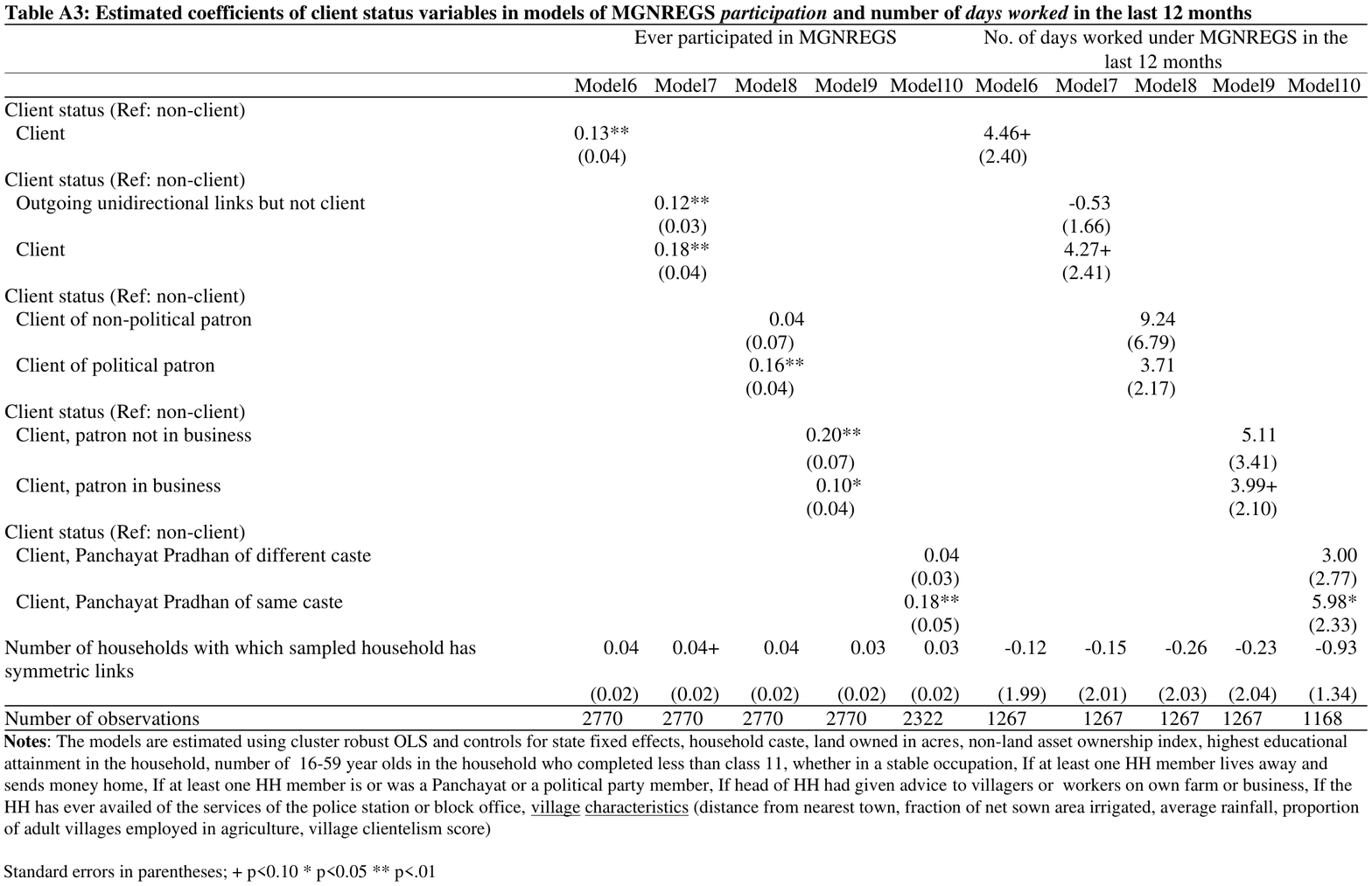}
\label{default}

\end{figure}

\newpage

\begin{figure}[htbp]

\includegraphics[angle=90, scale=0.7]{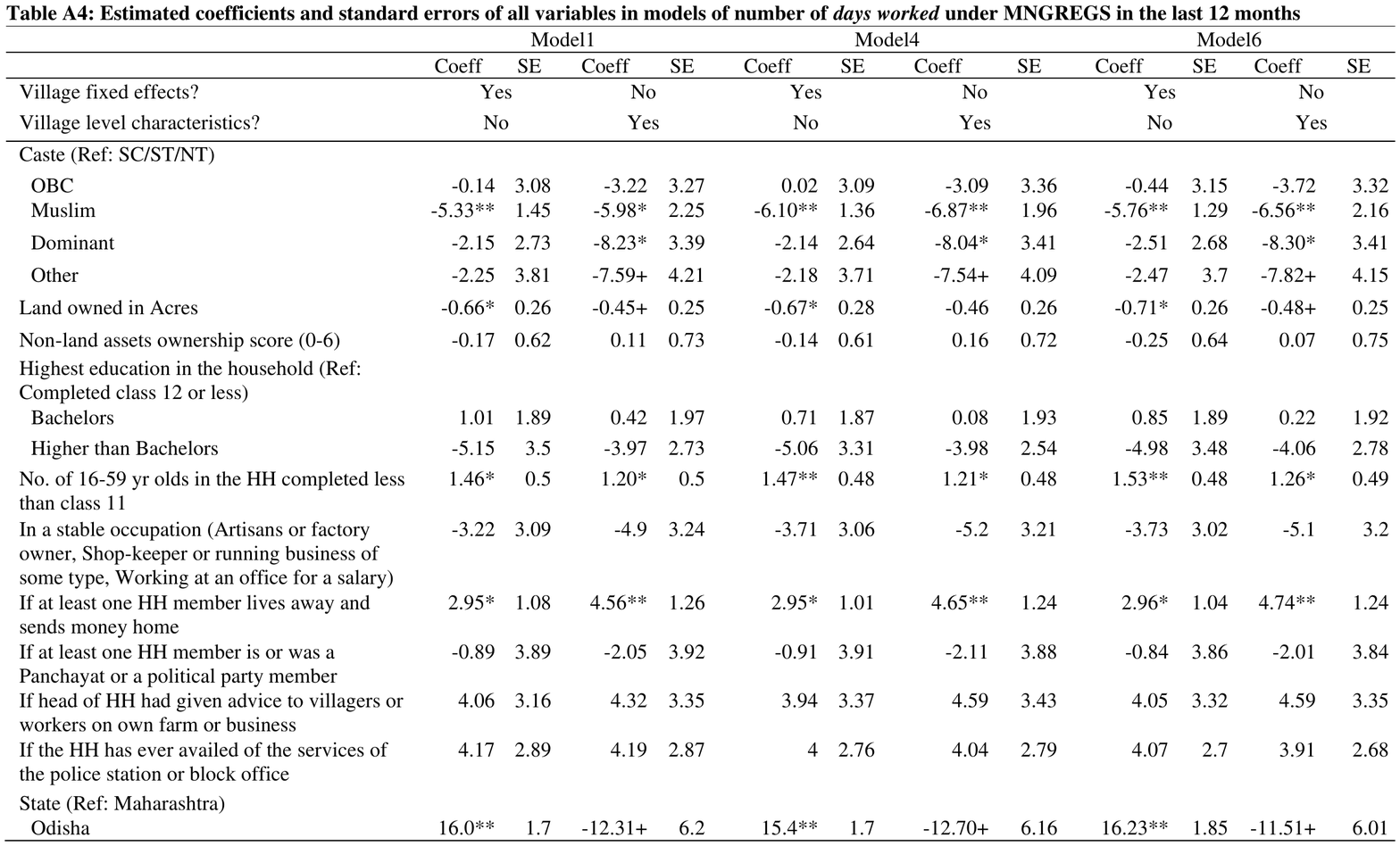}
\label{default}

\end{figure}

\newpage

\begin{figure}[htbp]

\includegraphics[angle=90, scale=0.7]{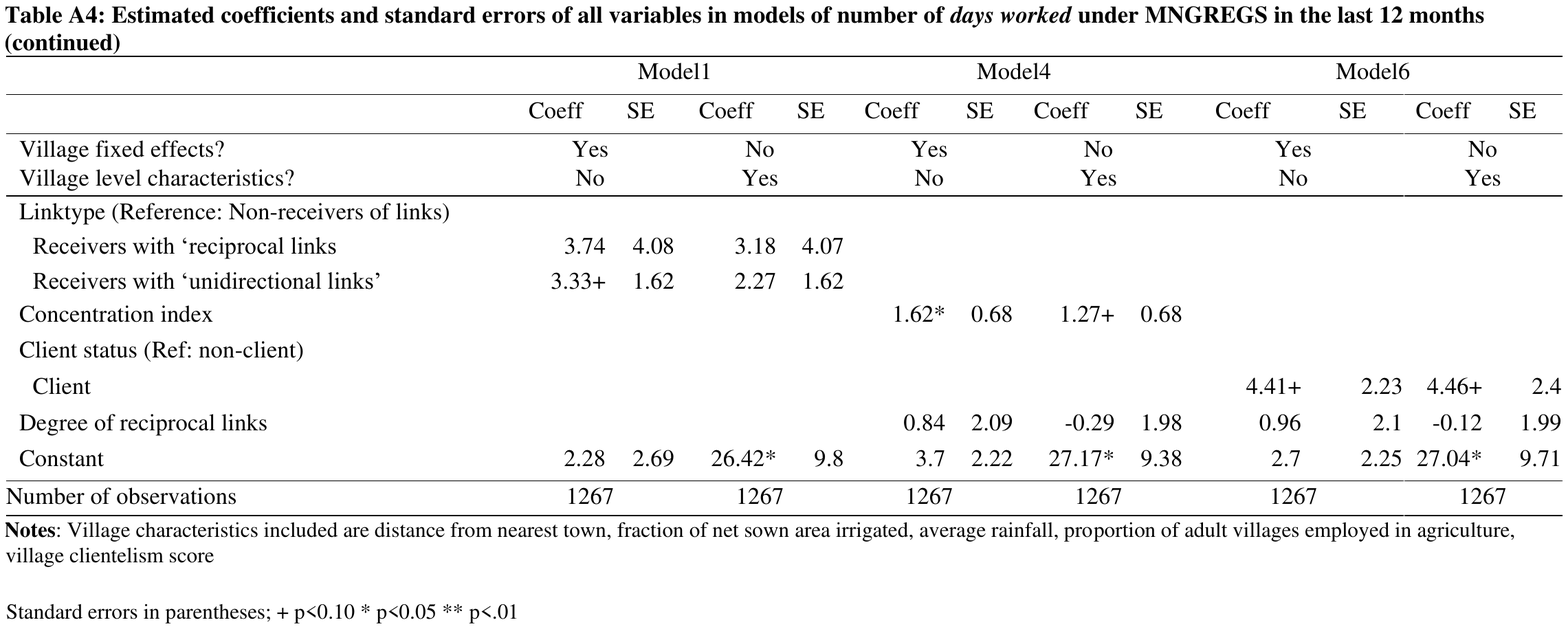}
\label{default}

\end{figure}

\newpage

\begin{figure}[htbp]

\includegraphics[angle=90, scale=0.7]{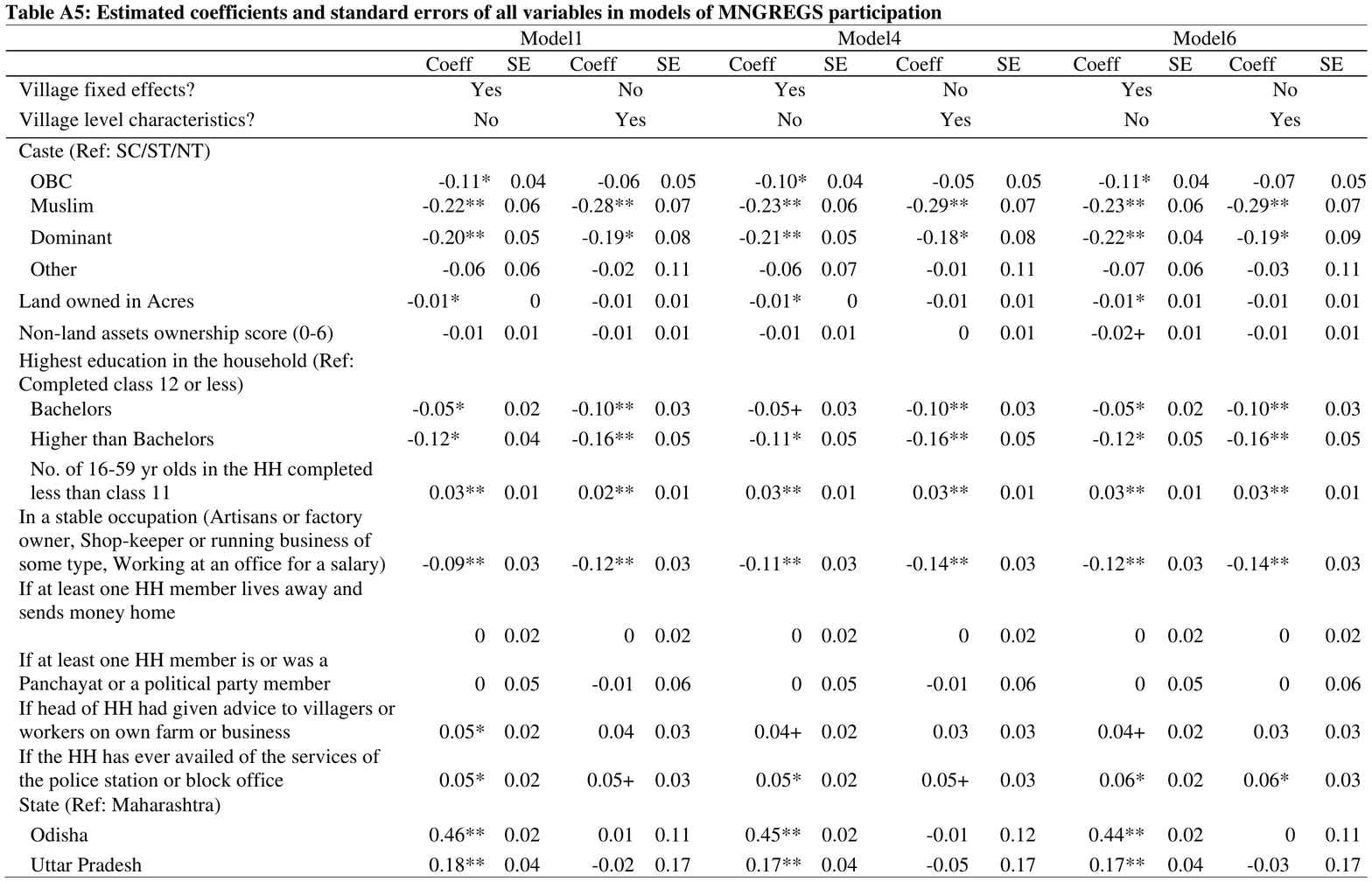}
\label{default}

\end{figure}

\newpage

\begin{figure}[htbp]

\includegraphics[angle=90, scale=0.7]{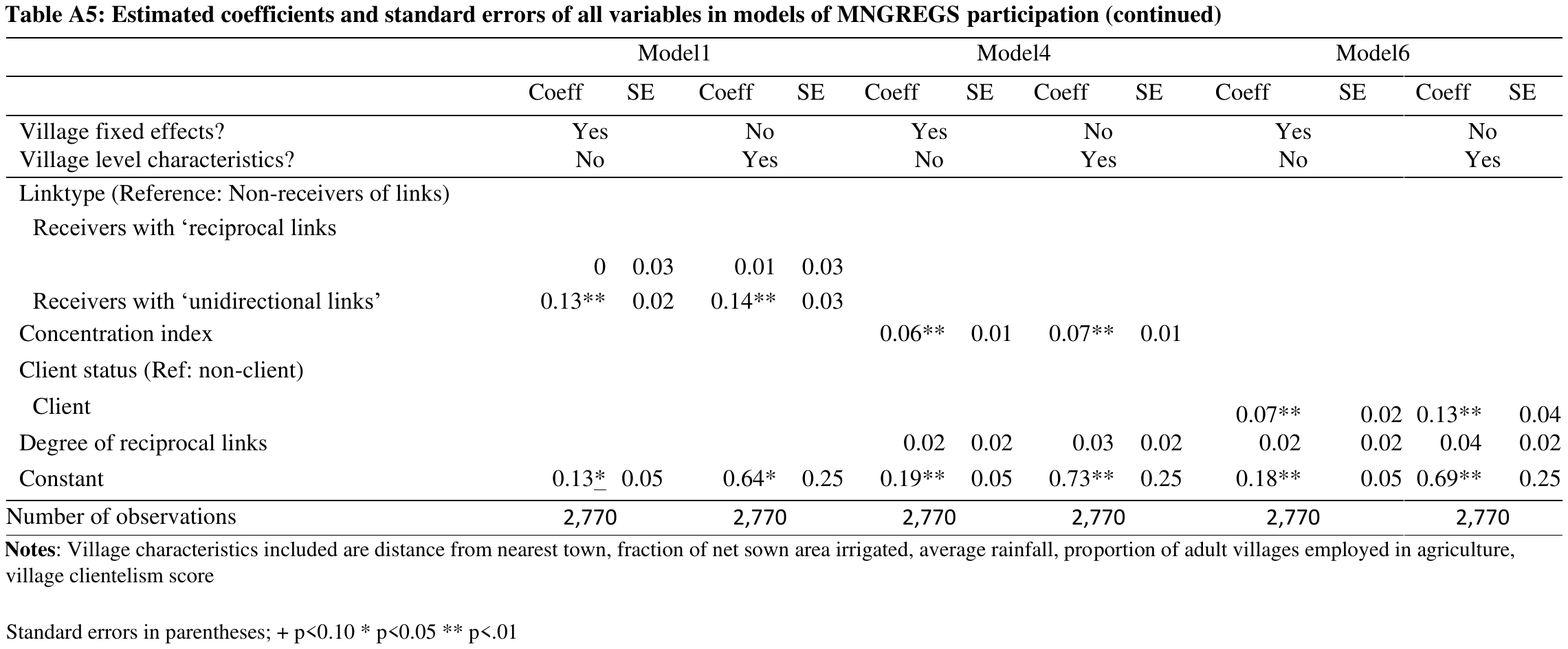}
\label{default}

\end{figure}

\end{document}